\documentclass{aa}
\usepackage{graphicx}
\bibpunct{(}{)}{;}{a}{}{,}
\usepackage{txfonts}
\usepackage{siunitx}
\usepackage{caption}
\usepackage{enumitem}
\usepackage{float}
\usepackage{placeins}
\usepackage{mwe,tikz}
\usepackage{changes}
\usepackage[colorlinks=true,linkcolor=blue,citecolor=blue,urlcolor=blue]{hyperref}
 
  \def\Hbeta{$\mathrm{H\beta}$}   
 
   \def\HeI{\ion{He}{i}}
 \def\HeII{\ion{He}{ii}}

   \def\OII{\ion{O}{ii}}
 \def\OIII{\ion{O}{iii}}  
 \def\OVI{\ion{O}{vi}}  
 \def\MgII{\ion{Mg}{ii}} 
 \def\AlIII{\ion{Al}{iii}}
 \def\SiII{\ion{Si}{ii}} \def\SiIII{\ion{Si}{iii}} \def\SiIV{\ion{Si}{iv}} 
 \def\NeII{\ion{Ne}{ii}} \def\NeIII{\ion{Ne}{iii}}

\newcommand{\msunpyr}{\ifmmode{\,M_{\odot}\,\mbox{yr}^{-1}} \else{ M$_{\odot}$/yr}\fi}

\newcommand{\kms}{\ifmmode{\,\mbox{km}\,\mbox{s}^{-1}}\else{km\,s^{-1}}\fi}
\newcommand{\kpc}{\ifmmode {\,\mbox{kpc}} \else{kpc}\fi}
\newcommand{\msun}{\ifmmode M_{\odot} \else M$_{\odot}$\fi}
\newcommand{\rsun}{\ifmmode R_{\odot} \else R$_{\odot}$\fi}
\newcommand{\lsun}{\ifmmode L_{\odot} \else L$_{\odot}$\fi}
\newcommand{\zsun}{\ifmmode Z_{\odot} \else $Z_{\odot}$\fi}
\newcommand{\xsun}{\ifmmode X_{\odot} \else $X_{\odot}$\fi}
\newcommand{\velo}{\ifmmode\varv\else$\varv$\fi}
\newcommand{\vinf}{\ifmmode\velo_\infty\else$\velo_\infty$\fi}
\newcommand{\rgal}{\ifmmode \,R_{\mathrm{gal}} \else R$_{\mathrm{gal}}$\fi}

\begin{document} 

    \title{Potsdam Wolf-Rayet stellar atmosphere grids of OB-type stars:}
    \subtitle{I. Spectroscopic temperature diagnostics across metallicity }
  
 	\titlerunning{}
 	
 	\author{D.~Pauli$^{\ref{inst1},\ref{inst4}}$  \and H.~Todt$^{\ref{inst1}}$ \and  W.-R.~Hamann$^{\ref{inst1}}$ \and L.~M.~Oskinova$^{\ref{inst1}}$ \and A.~A.~C.~Sander$^{\ref{inst2},\ref{inst:iwr},\ref{inst:cau}}$ \and T.~Shenar$^{\ref{inst3}}$ \and V.~Ramachandran$^{\ref{inst2}}$ \and S.~Reyero~Serantes$^{\ref{inst1}}$}
 	
 	\authorrunning{D. Pauli et al.}
 	
 	\institute{Institut f{\"u}r Physik und Astronomie, Universit{\"a}t Potsdam, Karl-Liebknecht-Str. 24/25, 14476 Potsdam, Germany\label{inst1} 
 	\and Institute of Astronomy, KU Leuven, Celestijnenlaan 200D, 3001 Leuven, Belgium\label{inst4}
    \and {Zentrum f{\"u}r Astronomie der Universit{\"a}t Heidelberg, Astronomisches Rechen-Institut, M{\"o}nchhofstr. 12-14, 69120 Heidelberg\label{inst2}} 
    \and {Universit\"at Heidelberg, Interdiszipli\"ares Zentrum f\"ur Wissenschaftliches Rechnen, 69120 Heidelberg, Germany\label{inst:iwr}}
    \and {Institut f{\"u}r Theoretische Physik und Astrophysik, Christian-Albrechts-Universit{\"a}t zu Kiel, Leibnizstr.\ 15, 24118 Kiel, Germany\label{inst:cau}}
    \and {The School of Physics and Astronomy, Tel Aviv University, Tel Aviv 6997801, Israel\label{inst3}}}
 	
 	\date{Received ; Accepted}
 	
 	\abstract{
        Massive hot stars are among the major ionizing sources in the Universe. The ionizing flux of a star strongly depends on its temperature, which is best measured spectroscopically using lines of different ionization stages of the same element. For the quantitative spectral analysis of O-type stars, helium lines are commonly employed, and for B-type stars, lines of silicon and magnesium are used. However, at low metallicity, many of the diagnostic metal lines disappear.
    }
 	  {    
        Here, we conduct a systematic theoretical analysis of stellar atmosphere models of OB stars to study the effect of metallicity on the behavior of key temperature diagnostic lines and how this impacts their spectral classification.
    }
    {
        We computed large grids of state-of-the-art non-local thermal equilibrium (non-LTE) stellar atmosphere models at four different metallicities, ranging from $\zsun$ (i.e., solar) to $1/31\,\zsun$. We presented contour plots for the equivalent widths of selected diagnostic lines in the temperature-gravity plane and investigated their dependence on metallicity. Using Galactic stellar templates of O- and B-type stars, we further established equivalent width ratios for spectral classification and applied them to our models.
    }
    {
        For the hottest stars in our model grids, \HeI{} lines become detectable at lower metallicity due to the reduced back-warming in the atmosphere, opening new possibilities for their classification. Metal lines used for the analysis of B-type stars are mostly absent in the spectra of stars with metallicity below that of the Small Magellanic Cloud, except \SiIII{}\,$\lambda4553$ and \SiIV{}\,$\lambda4089$, challenging current classification schemes. 
    }
    {
        Our stellar atmosphere grids reveal how metallicity affects key temperature diagnostics. Due to reduced back-warming and non-LTE effects in denser stellar atmospheres, these diagnostic lines appear at higher stellar temperatures in lower-metallicity environments. However, the decrease in metallicity also causes many metal lines to disappear, leaving different sensitivities of \HeI{} lines the only viable temperature indicators for B-type stars in the optical. The ionizing flux of OB-type stars can be partially trapped within the stellar wind. As a result, at lower metallicity, a larger fraction of stars is able to emit hard ionizing radiation.
    }
 	
 	\maketitle
    \nolinenumbers
  
 	\section{Introduction} 
 	\label{sec:intro} 

        Massive stars ($M_\mathrm{ini}>8\,\mathrm{M}_\odot$) are the powerful cosmic engines that have shaped the Universe since the dawn of time. Due to their high temperatures, these stars emit copious amounts of UV radiation, which is used to ionize hydrogen. This makes massive stars pivotal for the reionization of the Universe, as well as for the regulation of star formation in entire galaxies \citep{bar3:06, hop1:14}. Furthermore, these stars are responsible for the synthesis of heavy elements and their distribution to the interstellar medium (ISM) via their strong stellar winds and supernova explosions. 

        The flux of ionizing radiation (mostly UV) emitted by a star is mainly set by a star's surface temperature and its luminosity. Thus, to estimate the UV flux originating from massive stars, one needs to measure their temperatures. The atmospheric properties of a star, namely temperature, surface gravity, and chemical abundances, dictate the strengths of spectral lines present in a stellar spectrum. For OB stars specifically, quantitative analyses using equivalent widths (EWs) of specific lines have been extensively used to determine their temperature  \citep[e.g.,][]{sto1:68, con1:71, did1:82}.
        In spectroscopy analyses of O- and B-type stars, often ratios of elements of the same ionization stage, such as \HeI{} and \HeII{} are employed to estimate the temperature. However, these lines become insensitive to temperature for stars with $T\lesssim\SI{30}{kK}$ and $T\gtrsim\SI{45}{kK}$ as the ionization stages of \HeI{} and \HeII{}, respectively, cease to be populated. In those cases, metal lines are utilized as an alternative temperature diagnostic \citep{cro1:02,wal1:04,eva1:15,mce1:15}. Unfortunately, these diagnostics weaken or even vanish with decreasing metallicity. 
        As the availability of observed spectra of metal-poor OB stars is increasing, it is important to investigate which diagnostic lines are expected to be present in the spectra of these stars.
        
        Studies of massive stars in nearby low-metallicity galaxies are used to derive their stellar and feedback properties and are then used as input into spectral synthesis codes used to model the spectra of distant galaxies. One of the largest programs focusing on creating a spectral library of massive stars at low metallicities is the ULLYSES\footnote{\url{https://ullyses.stsci.edu/}} initiative \citep{rom1:20} and the XShootU\footnote{\url{https://massivestars.org/xshootu}} collaboration \citep{vin2:23}. Their aim is to provide a large representative sample of UV and optical spectra of OB-type stars in the Small and Large Magellanic Clouds (SMC and LMC, respectively) as well as selected objects in the low-metallicity galaxies Sextans~A and NGC~3109, probing a large metallicity range from $1/2\,\zsun$ to $Z\sim1/10\,\zsun$. 

        Even lower metallicities can be probed in local group galaxies, such as Sextans A with $Z\sim1/16\,\zsun$ \citep[e.g.,][]{tel1:24,fur1:25,gul1:26}, Leo A with ${Z\sim1/20\,\zsun}$ \citep[e.g.,][]{gul1:22}, or I\,Zw\,18 with ${Z\sim1/30\,\zsun}$ \citep{sea1:72,lec1:04}. Most of the extremely metal-poor galaxies are too far away for resolving individual stars with the current generation of telescopes and preventing large surveys needed to perform a detailed spectral analysis of their stellar content. However, to correctly model spectra of unresolved stellar populations, new model low-metallicity spectral grids are needed.
        
        \begin{table*}[tbp]
         	\centering
         	\caption{Chemical abundances used for the different grids.}
         	\begin{tabular}{c|cc|cc|cc|cc}\hline \hline \rule{0cm}{2.2ex}%
                Element & \multicolumn{2}{c|}{GAL}& \multicolumn{2}{c|}{LMC}& \multicolumn{2}{c|}{SMC}& \multicolumn{2}{c}{subSMC}\\
         		        & Abundances& Refs.& Abundances& Refs.&Abundances& Refs.&Abundances& Refs.\\
         		        & [mass frac.]&& [mass frac.]&& [mass frac.]&&[mass frac.]&\\
         		\hline \rule{0cm}{2.8ex}%
         		\rule{0cm}{2.4ex}H  & $0.7375$          & Asp & $0.7375$               & Asp             & $0.7375$               & Asp             & $0.7375$               & Asp \\
         		\rule{0cm}{2.4ex}He & $0.2508$          & Equ & $0.2579$               & Equ             & $0.2605$               & Equ             &  $0.2621$              & Equ \\
         		\rule{0cm}{2.4ex}C  & $\num{236e-5}$    & Asp &\,\,\,$\num{48e-5}$     & Hun, Tru             &\,\,\,$\num{21e-5}$     & Hun, Tru        &\,\,\,\,\,\,$\num{7e-5}$& $1/31\cdot$\,Asp \\
         		\rule{0cm}{2.4ex}N  &\,\,\,$\num{69e-5}$& Asp &\,\,\,\,\,\,$\num{8e-5}$& Hun, Tru        &\,\,\,\,\,\,$\num{3e-5}$& Hun, Tru        &\,\,\,\,\,\,$\num{2e-5}$& $1/31\cdot$\,Asp \\
         		\rule{0cm}{2.4ex}O  & $\num{573e-5}$    & Asp & $\num{264e-5}$         & Hun, Tru        & $\num{113e-5}$         & Hun, Tru        &\,\,\,$\num{18e-5}$     & $1/31\cdot$\,Asp \\
         		\rule{0cm}{2.4ex}Ne & $\num{125e-05}$   & Asp &\,\,\,$\num{63e-5}$     & $1/2\cdot$\,Asp & \,\,\,$\num{18e-5}$    & $1/7\cdot$\,Asp &\,\,\,\,\,\,$\num{4e-5}$& $1/31\cdot$\,Asp \\
         		\rule{0cm}{2.4ex}Mg &\,\,\,$\num{69e-5}$& Sco &\,\,\,$\num{21e-5}$     & Hun, Tru        & \,\,\,$\num{10e-5}$    & Hun, Tru        &\,\,\,\,\,\,$\num{2e-5}$& $1/31\cdot$\,Sco \\
         		\rule{0cm}{2.4ex}Al &\,\,\,$\num{53e-6}$& Sco &\,\,\,$\num{27e-6}$     & $1/2\cdot$\,Sco & \,\,\,$\num{8e-6}$     & $1/7\cdot$\,Sco &\,\,\,\,\,\,$\num{2e-6}$& $1/31\cdot$\,Sco \\
         		\rule{0cm}{2.4ex}Si &\,\,\,$\num{67e-5}$& Sco &\,\,\,$\num{32e-5}$     & Hun, Tru        & \,\,\,$\num{13e-5}$    & Hun, Tru        &\,\,\,\,\,\,$\num{2e-5}$& $1/31\cdot$\,Sco \\
         		\rule{0cm}{2.4ex}P  &\,\,\,$\num{58e-7}$& Sco &\,\,\,$\num{29e-7}$     & $1/2\cdot$\,Sco &\,\,\,\,\,\,$\num{8e-7}$& $1/7\cdot$\,Sco &\,\,\,\,\,\,$\num{2e-7}$& $1/31\cdot$\,Sco \\
         		\rule{0cm}{2.4ex}S  &\,\,\,$\num{31e-5}$& Sco &\,\,\,$\num{16e-5}$     & $1/2\cdot$\,Sco &\,\,\,\,\,\,$\num{4e-5}$& $1/7\cdot$\,Sco &\,\,\,\,\,\,$\num{1e-5}$& $1/31\cdot$\,Sco \\
         		\rule{0cm}{2.4ex}G  & $\num{130e-5}$    & Sco &\,\,\,$\num{70e-5}$     & Tru             & \,\,\,$\num{35e-5}$    & Tru             &\,\,\,\,\,\,$\num{4e-5}$& $1/31\cdot$\,Sco\\
         		\hline
         	\end{tabular}
         	\label{tab:abundances}
            \begin{minipage}{0.95\linewidth}
                \ignorespaces 
                \rule{0cm}{2.4ex}\textbf{References.} Asp: \citet{asp1:05}; Equ: $Y=1-X-Z$; Sco: \citet{sco2:15}; Hun: \citet{hun1:07}; Tru: \citet{tru1:07}
            \end{minipage}
         \end{table*}  

        While numerous stellar atmosphere model grids exist in the literature, spanning metallicities from $\zsun$ down to $1/30\,\zsun$ \citep[e.g.,][]{lan1:03,lan1:07,hai1:19,mar1:21,mar1:24}, none covers the full range in metallicity with a homogeneous set of models. As a result, different grids must often be combined, despite relying on varying assumptions for covered stellar and wind parameters. In some cases, different atmosphere codes accounting for different physics within a stellar atmosphere are used, which introduces additional systematic uncertainties.
        
        In this paper, we present new large grids of non-LTE stellar atmosphere models of OB stars covering metallicities from $\zsun$ to $1/31\,\zsun$ and using adequate mass-loss rates. We utilize our model grid to calculate the EWs of key temperature diagnostic lines to obtain a theoretical understanding of the spectral appearance of massive stars as a function of metallicity. 

        The paper is structured as follows. In Sect.\,\ref{sec:methods}, we outline the choices made within our stellar atmosphere models, and the criteria used to set stellar and wind properties for the grid models. Sect.\,\ref{sec:products} describes the data products provided by our new grids and how these products can be accessed. Our new results on the temperature diagnostics across the various metallicities are presented in Sect.\,\ref{sec:results}. The implications and limitations of our results are discussed in Sect.\,\ref{sec:discuss}. In Section~\ref{sec:conclusions}, we summarize the predictions from our stellar atmosphere grids.

 	\section{Methods}
 	\label{sec:methods}

    \subsection{Input parameters for the stellar atmosphere models}
    \label{sec:grid}
    \citet{hai1:19} presented the first generation of Potsdam Wolf-Rayet (PoWR) stellar atmosphere model grids for OB-type stars at solar, LMC, and SMC metallicities. However, these grids include only stars with initial masses up to $60\,\msun$, limiting their applicability to the most massive stellar populations, which are also the brightest and therefore most readily observed in distant galaxies. Furthermore, the solar and LMC models were computed assuming fixed mass-loss rates, while the SMC grids were parameterized by a fixed wind-strength parameter, $\log Q$. This approach is suboptimal for spectral population synthesis applications and may introduce biases in the inferred spectral types of stars with strong winds.

    Beyond the underlying physical assumptions, the PoWR code itself has undergone significant developments in recent years. These improvements include an updated temperature correction scheme, enhanced numerical treatment of the comoving-frame (CMF) radiative transfer, and a refactored spectral synthesis module. In particular, the treatment of the Eddington factors and the excitation temperature for the iron group super levels turned out to be crucial. In addition, the atomic data have been expanded and updated impacting not only the spectral appearance but also the line driving of the wind. Moreover, several several inaccuracies affecting individual line transitions. The model outputs have also been extended to include additional diagnostics, such as synthetic photometry (e.g., for Gaia), as well as a continuous wavelength coverage from the far-UV to the infrared, spanning the range accessible to facilities such as JWST. This broad wavelength coverage was not available in any previously published public model grid.
    
    \subsubsection{Fundamental stellar parameters}
    \label{sec:fundamental_parameters}
    
        In the approximation of a plane-parallel, static stellar atmosphere, two main parameters define the emergent spectrum of a star: its temperature, $T$, and surface gravity, $\log\,g$. In this case, the stellar (bolometric) luminosity, $L$, has only the effect of a scaling factor. To create a widely applicable grid of OB star spectra, we have structured our grid to encompass temperatures ranging from ${T=\SIrange{15}{62}{kK}}$ in steps of ${\Delta T=\SI{1}{kK}}$ and surface gravities within the range of ${\log(g/(\si{cm\,s^{-2}}))=\SIrange{4.4}{2.0}{}}$, in intervals of ${\Delta \log(g/(\si{cm\,s^{-2}}))=0.2}$.
        
        Since our atmosphere models account for spherical geometry and expansion (i.e., stellar wind), more model parameters are required.
        In the case of a spherically extended stellar atmosphere, the ``stellar radius'' becomes a matter of definition. In our model calculations, the inner boundary, namely the stellar radius, $R_\ast$, is defined at a Rosseland mean continuum optical depth of $\tau_\mathrm{Ross}=20$.
        
        Using luminosity, $L$, as an input parameter of a model, the above definition of $R_\ast$ leads to a corresponding definition of the stellar temperature, $T_\ast$, via the Stefan-Boltzmann relation 
        \begin{equation}
        L = 4 \pi~R_\ast^2~ \sigma_\mathsf{SB} T_\ast^4\ ,
        \end{equation}
        where $\sigma_\mathsf{SB}$ is the Stefan-Boltzmann constant. 
        Alternatively, one might use the mean optical depth of 2/3 for the reference radius. For OB star models with optically thin winds, the corresponding effective temperature $T_{\rm eff}$ is marginally lower ($\lesssim1\%$) than $T_\ast$. 

        The purpose of these grids is to create stellar atmosphere models with properties similar to those of real stars. The previous model grids from \citet{hai1:19} only extended up to $60\,\msun$, limiting their application to the most massive stars in a population. To overcome this limitations, we employ the nonrotating single-star MIST stellar evolutionary tracks \citep{cho1:16} that extend up to $150\,\msun$ stars. Provided that the evolutionary tracks of OB stars do not crossover, one can assign a unique luminosity $L$ to any combination of $T$ and $\log g$ covered by our grid. Note that the MIST models report a $T_\mathrm{eff,\,MESA}$ based on a gray atmosphere. However, since for stars with optical thin winds the differences between the different temperature notations are very small ($\lesssim1\%$), we use these quantities as equivalents. Since stellar evolution models have a finite resolution, a given pair of $T$ and $\log g$ might lie between two tracks. In this case, we perform an interpolation (see Appendix~\ref{app:interpol}). Figures~\ref{fig:HRD} and \ref{fig:kiel}
        show the HRDs and the $T_\ast$ -- $\log g$ (Kiel) diagrams correspondingly, with the MIST evolutionary tracks at metallicities of $[\mathrm{Fe}/\mathrm{H}] = 0.0$ (GAL), $[\mathrm{Fe}/\mathrm{H}] = -0.5$ (LMC), $[\mathrm{Fe}/\mathrm{H}] = -1.0$ (SMC), and $[\mathrm{Fe}/\mathrm{H}] = -1.5$ (sub-SMC) overlayed by squares indicating the parameters of the models calculated for the respective grid.

        \begin{figure*}[thb]
            \centering
            \includegraphics[trim={3.75cm 3.5cm 4cm 4cm},clip, width=0.95\textwidth]{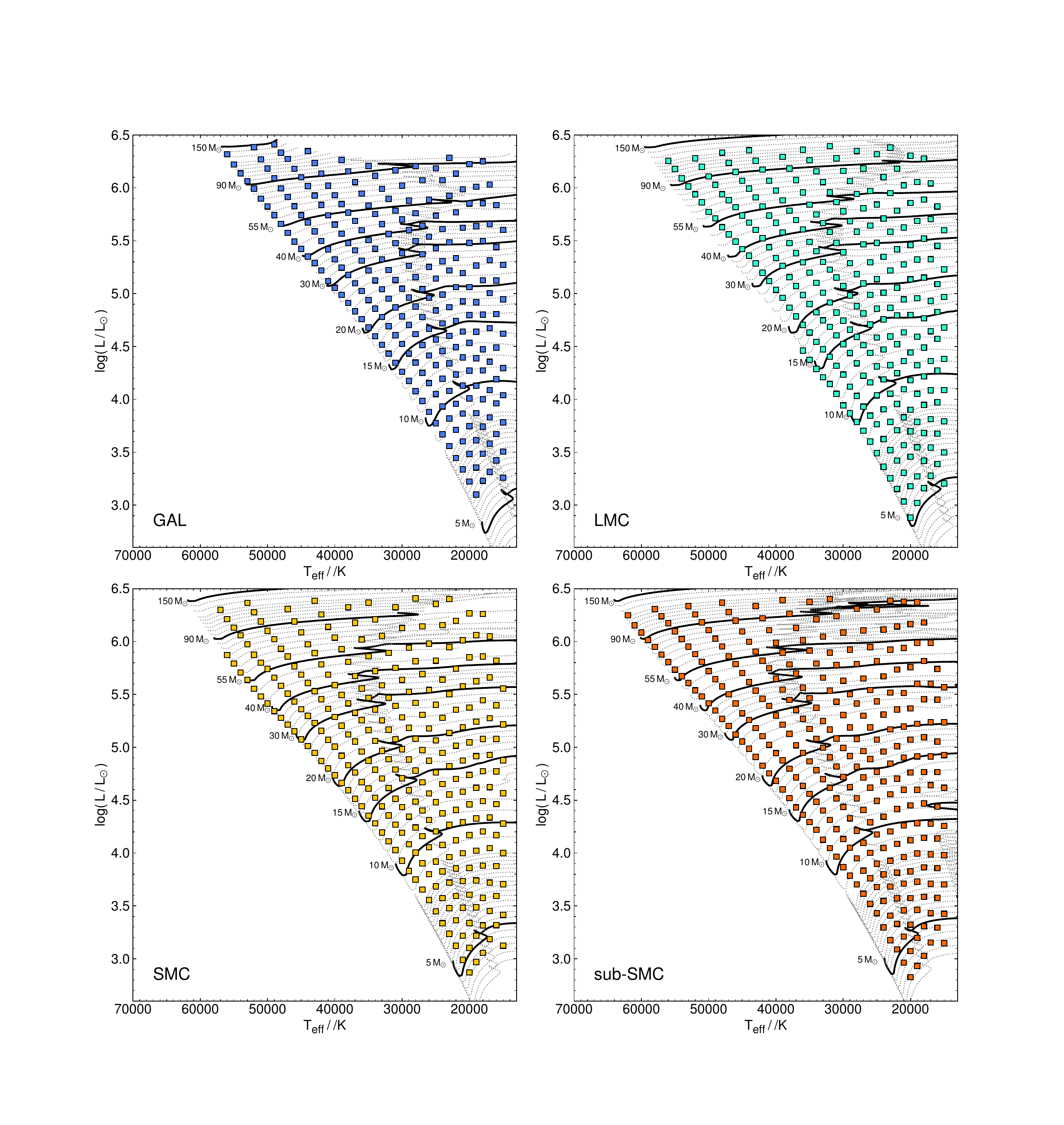}
            \caption{HRDs showing the MIST evolutionary tracks (lines) over-plotted by the positions covered by the computed stellar atmosphere model grids (squares). The MIST tracks for solar metallicity (GAL), LMC, SMC and sub-SMC metallicity are shown in the upper left, upper right, lower left, and lower right panels, respectively. All available MIST tracks within the initial mass range of $M=\numrange{4}{150}\,\msun$ are illustrated as lines. For clarity we show only the tracks with initial masses $5\,\msun$, $10\,\msun$, $15\,\msun$, $20\,\msun$, $30\,\msun$, $40\,\msun$, $55\,\msun$, $90\,\msun$, and $150\,\msun$ as solid black lines and the remaining tracks as dotted gray lines. We show here the $T_\mathrm{eff}$ from the MESA models and the $T_\ast$ of the stellar atmosphere models. Note that for OB stars that have optically thin winds, the difference between $T_\mathrm{eff}$ and $T_\ast$ is very small ($\lesssim1\%$).}
            \label{fig:HRD}
        \end{figure*}
        
    \subsubsection{Surface abundances}
        
        In addition to the fundamental stellar parameters, the element abundances have a noticeable effect on stellar spectra.
        Our models only cover the early evolutionary phases of hot massive stars ranging from the main sequence ($\log(g/(\mathrm{cm\,s^{-2}}))=4.4$) to the supergiant stage ($\log(g/(\mathrm{cm\,s^{-2}}))=2.0$). For simplicity, we assume fixed abundances for all models at a given metallcity.
        
        Within our stellar atmosphere models, detailed model atoms for the elements of H, He, C, N, O, Ne, Mg, Al, Si, P, and S are included. In addition to \citet{hai1:19}, we also account for Ne and Al, as lines from these elements are detectable in spectra of Galactic massive stars and provide additional opacities that are important for, for example line driving of stellar winds.

        Iron and the iron-group elements (Fe, Sc, Ti, V, Cr, Mn, Fe, Co, and Ni) are extremely important for the effects of line blanketing and back-warming. Due to their many electrons, they have millions of energy levels and billions of line transitions, which makes an exact non-LTE treatment computationally impossible. Since their ionization energies are similar, one can combine the model atoms of these iron-group elements to one generic element ``G''. Moreover, this element is treated in the ``superlevel approximation'': the millions of energy levels are combined to a much smaller number of energy bands, where all energy levels within such a ``superlevel'' and with same parity share the same departure from LTE \citep[for details see][]{gra1:02}.
              
        For our Galactic grid (GAL; $\zsun=0.0012$), we adopt solar abundances as compiled by \citet{asp1:09}, and the updated values for heavier elements from \citet{sco2:15}. For our LMC ($Z_\mathrm{LMC}=1/2\,\zsun$; \citealt{rol1:02}), SMC ($Z_\mathrm{SMC}=1/7\,\zsun$; \citealt{hun1:07,tru1:07}), and sub-SMC ($Z_\mathrm{subSMC}=1/31\,\zsun$, i.e., $[\mathrm{Fe}/\mathrm{H}] = -1.5$) we scaled most of the abundances relative to the solar abundances of the GAL grid. Only for the C, N, and O elements in the LMC and SMC models, tailored abundances are used \citep{hun1:07,tru1:07}. A detailed list of all elements and their abundances is provided in Table~\ref{tab:abundances}.

    \subsubsection{Wind parameters}

        The stellar wind has a dominant impact on the UV spectra, where it can manifest itself in resonance lines of abundant ions. Those lines often show up as so-called P\,Cygni profiles, a blue-shifted absorption part accompanied by a red-shifted emission feature. The strength of the P\,Cygni profiles depends on the mass-loss rate, $\dot{M}$, and its width is set by the terminal wind velocity, $\varv_\infty$. The detailed shape of those profiles is influenced by the change of the wind velocity with radius, namely the velocity law (Sect.~\ref{sec:PoWR}). 

        The PoWR stellar atmosphere model grids for OB stars that we provide on the PoWR webpage\footnote{\url{www.astro.physik.uni-potsdam.de/PoWR/}} come in two flavors:    
        \begin{enumerate}        
            \item  Grids where $\dot{M}$ (or a related parameter) fixed for all models. In a few cases, grids with different sets of $\dot{M}$ are available \citep[][]{hai1:19}. Those grids are apt for the analysis of observations, either to obtain a first guess before calculating customized models, or to be applied as a database for automated spectral fitting or emulators. 
        
            \item Grids where $\dot{M}$ of each model is individually specified using a mass-loss recipe as function of the fundamental stellar parameters (this work). Such grids are better suited for spectral population synthesis, but can also be used as starting points for a spectral analysis.    
        \end{enumerate}
        
        The present paper focuses on the second type of model grids, for which a prescription of $\dot{M}$ is required. Various empirical mass-loss recipes have been given in the recent literature. Naturally, they are based on observed samples of stars and
        thus, established only for a limited parameter range. Therefore, for our grid calculations, we prefer to employ a theoretical mass-loss recipe that has a broader applicability range. The most commonly used mass-loss prescription for OB-type stars is the one by \citet{vin1:01}. This recipe is also used in the MIST stellar evolution tracks \citep{cho1:16}, which we employ to obtain the luminosity for each model to be calculated (cf.\ Sect.\,\ref{sec:fundamental_parameters}). 
        
\begin{figure}[tb]
            \centering
            \includegraphics[width=0.5\textwidth]{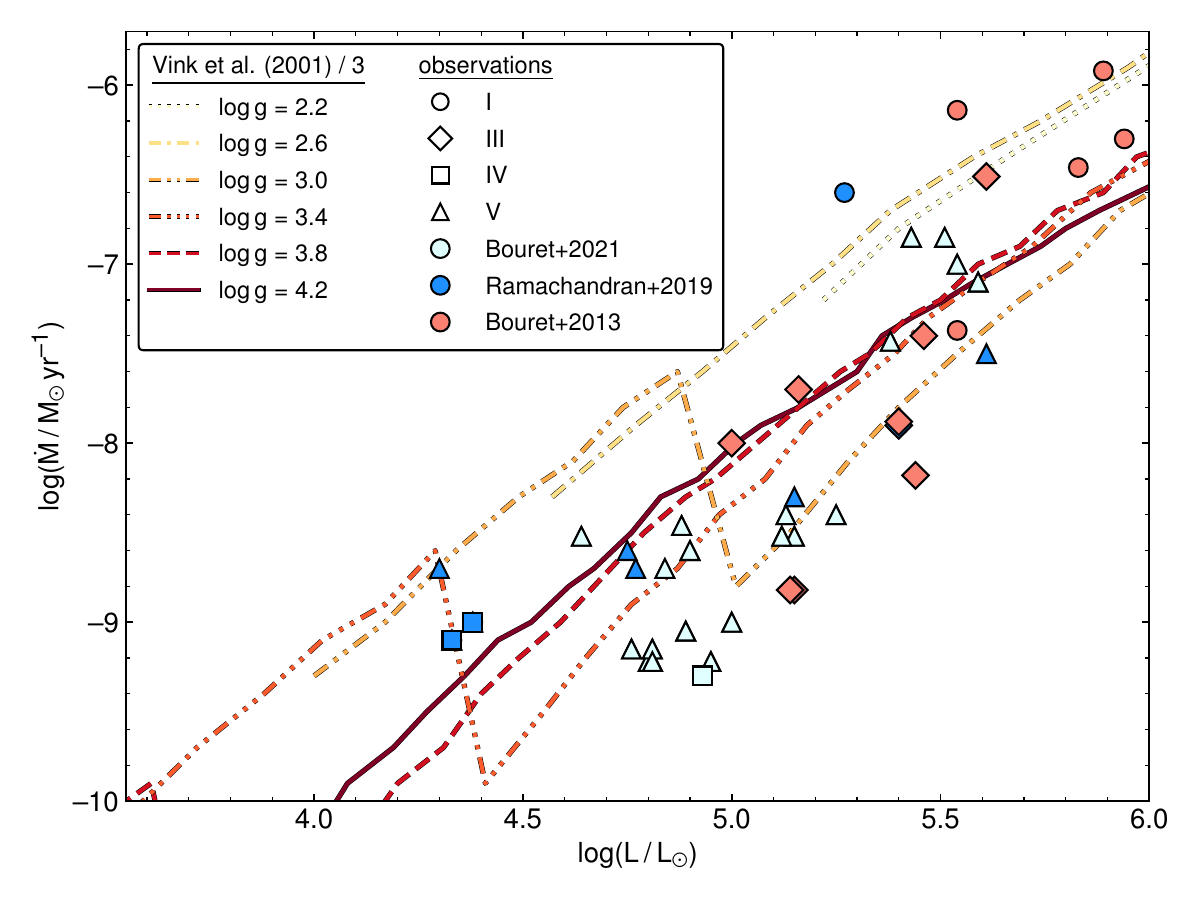}
            \caption{\citet{vin1:01} mass-loss rates divided by a factor of three (lines) compared to empirical measurements from OB-type stars in the SMC (symbols): The theoretical mass-loss rates are shown for different values of $\log\,g$ (see legend). The mass-loss rates of the OB-type stars are adopted from \citet{ram1:19} and \citet{bou1:13,bou1:21}. Their luminosity classes are indicated by different symbols.}
            \label{fig:mdot}
        \end{figure}

        However, there is a systematic discrepancy between Vink's prediction and empirical mass-loss rates. A whole conference dealt with this issue and concluded that empirical $\dot{M}$ values are typically three times lower than theoretically predicted, which can be attributed to wind clumping \citep{ham3:08}. This factor has been corroborated by various studies of Galactic OB-type stars
        \citep[e.g.,][]{ful1:06,osk1:07,kob1:19}.  For lower metallicity, studies also indicate such a discrepancy \citep[e.g.,][]{ram1:19, ric1:22}.  However, no offset factor is determined. We collected data for OB-type stars in the SMC and compared the mass-loss
        rates with the theoretical prediction by  \citet{vin1:01}. Only after the theoretical $\dot{M}$ values have been divided by three, the observed mass-loss rates can be matched (see Fig.~\ref{fig:mdot}). Since the offset of a factor of three seems to be metallicity independent,  we decided to adopt in our models mass-loss rates from \citet{vin1:01} divided by a factor of three.
        
        The theory of radiation-driven winds by \citet{cas1:75} and its later extensions \citep[e.g.,][]{fri1:86,pau1:86,kud1:89} predict that the terminal wind velocity, $\varv_\infty$, is proportional to the (effective) escape velocity, $\varv_\mathrm{esc,\,\Gamma}$,
        \begin{equation}
            \label{eq:vinfvesc}
            \varv_\infty = n\cdot\varv_\mathrm{esc,\,\Gamma} = n\cdot\sqrt{\frac{2GM}{R}(1-\Gamma_\mathrm{e})}.
        \end{equation}
        The quantity $\Gamma_\mathrm{e}$ denotes the Eddington parameter ($\propto L/M$) only accounting for electron scattering. \citet{lam1:95} demonstrated that the scaling factor $n$ likely depends on the evolutionary stage. Following \citet{vin1:01}, we employ for the hotter stars of our grid ($T\gtrsim\SI{25}{kK}$) a relation of $\varv_\infty=2.6\,\varv_\mathrm{esc,\,\Gamma}$, while for the cooler stars of our grid ($T\lesssim\SI{25}{kK}$) a shallower relation of $\varv_\infty=1.3\,\varv_\mathrm{esc,\,\Gamma}$ is used.   
        
        From the principles of radiative driving, a metallicity dependence of $\varv_\infty$ and thus also of Eq.\,\eqref{eq:vinfvesc} is expected in the form $\varv_\infty \propto Z^\alpha$. However, as there is no clear consensus on the scaling and both the theoretically and empirically obtained exponents $\alpha$ are not large \citep[e.g.,][]{lei1:92,vin1:21,haw1:24, Bernini-Peron+2024}. Therefore, we do not apply a metallicity scaling to $\varv_\infty$ in our grids and use Eq.\,\eqref{eq:vinfvesc} for all grids presented in this work, which is in contrast to the previous model grids from \citet{hai1:19}.

    \subsection{Stellar atmosphere modeling}
    \label{sec:PoWR}

        \begin{table*}[tbp]
         	\centering
         	\caption{Ionization stages used in the models for the different temperature regimes.}
         	\begin{tabular}{c|c|c|c|c|c}\hline \hline \rule{0cm}{2.2ex}%
                Element            & $T<\SI{21.5}{kK}$                          & $\SI{21.5}{kK}<T<\SI{28.5}{kK}$             & $\SI{28.5}{kK}<T<\SI{33.5}{kK}$                       & $\SI{33.5}{kK}<T<\SI{38.5}{kK}$              & $\SI{38.5}{kK}<T$                           \\
                \hline
                \rule{0cm}{2.4ex}H & \sc{i}, \sc{ii}                            & \sc{i}, \sc{ii}                             & \sc{i}, \sc{ii}                                       & \sc{i}, \sc{ii}                              & \sc{i}, \sc{ii}                             \\
                \rule{0cm}{2.4ex}He& \sc{i}, \sc{ii}, \sc{iii}                  & \sc{i}, \sc{ii}, \sc{iii}                   & \sc{i}, \sc{ii}, \sc{iii}                             & \sc{i}, \sc{ii}, \sc{iii}                    & \sc{i}, \sc{ii}, \sc{iii}                   \\
                \rule{0cm}{2.4ex}C & \sc{i}, \sc{ii}, \sc{iii}, \sc{iv}, \sc{v} & \sc{i}, \sc{ii}, \sc{iii}, \sc{iv}, \sc{v}  & \sc{i}, \sc{ii}, \sc{iii}, \sc{iv}, \sc{v}            & \sc{ii}, \sc{iii}, \sc{iv}, \sc{v}           & \sc{iii}, \sc{iv}, \sc{vi}                  \\
                \rule{0cm}{2.4ex}N & \sc{i}, \sc{ii}, \sc{iii}, \sc{iv}         & \sc{i}, \sc{ii}, \sc{iii}, \sc{iv}, \sc{v}  & \sc{i}, \sc{ii}, \sc{iii}, \sc{iv}, \sc{v}            & \sc{ii}, \sc{iii}, \sc{iv}, \sc{v}, \sc{vi}  & \sc{iii}, \sc{iv}, \sc{v}, \sc{vi}          \\
                \rule{0cm}{2.4ex}O & \sc{i}, \sc{ii}, \sc{iii}, \sc{iv}         & \sc{i}, \sc{ii}, \sc{iii}, \sc{iv}, \sc{v}  & \sc{i}, \sc{ii}, \sc{iii}, \sc{iv}, \sc{v}            & \sc{ii}, \sc{iii}, \sc{iv}, \sc{v}, \sc{vi}  & \sc{ii}, \sc{iii}, \sc{iv}, \sc{v}, \sc{vi} \\
                \rule{0cm}{2.4ex}Ne& \sc{i}, \sc{ii}, \sc{iii}                  & \sc{i}, \sc{ii}, \sc{iii}, \sc{iv}          & \sc{ii}, \sc{iii}, \sc{iv}                            & \sc{ii}, \sc{iii}, \sc{iv}, \sc{v}           & \sc{ii}, \sc{iii}, \sc{iv}, \sc{v}, \sc{vi} \\
                \rule{0cm}{2.4ex}Mg& \sc{ii}, \sc{iii}                          & \sc{ii}, \sc{iii}                           & \sc{ii}, \sc{iii}, \sc{iv}                            & \sc{ii}, \sc{iii}, \sc{iv}                   & \sc{ii}, \sc{iii}, \sc{iv}, \sc{v}          \\
                \rule{0cm}{2.4ex}Al& \sc{ii}, \sc{iii}, \sc{iv}                 & \sc{iii}, \sc{iv}                           & \sc{iii}, \sc{iv}                                     & \sc{iii}, \sc{iv}, \sc{v}                    & \sc{iii}, \sc{iv}, \sc{v}                   \\
                \rule{0cm}{2.4ex}Si& \sc{ii}, \sc{iii}, \sc{iv}, \sc{v}         & \sc{ii}, \sc{iii}, \sc{iv}, \sc{v}          & \sc{ii}, \sc{iii}, \sc{iv}, \sc{v}, \sc{vi}           & \sc{ii}, \sc{iii}, \sc{iv}, \sc{v}, \sc{vi}  & \sc{iii}, \sc{iv}, \sc{v}, \sc{vi}, \sc{vii}\\
                \rule{0cm}{2.4ex}P & \sc{iii}, \sc{iv}, \sc{v}                  & \sc{iii}, \sc{iv}, \sc{v}, \sc{vi}          & \sc{iii}, \sc{iv}, \sc{v}, \sc{vi}                    & \sc{iv}, \sc{v}, \sc{vi}                     & \sc{iv}, \sc{v}, \sc{vi}, \sc{vii}          \\
                \rule{0cm}{2.4ex}S & \sc{ii}, \sc{iii}, \sc{iv}, \sc{v}         & \sc{ii}, \sc{iii}, \sc{iv}, \sc{v}, \sc{vi} & \sc{ii}, \sc{iii}, \sc{iv}, \sc{v}, \sc{vi}, \sc{vii} & \sc{iii}, \sc{iv}, \sc{v}, \sc{vi}, \sc{vii} & \sc{iv}, \sc{v}, \sc{vi}, \sc{vii}          \\
                \rule{0cm}{2.4ex}G & \sc{i}, \sc{ii}, \sc{iii}, \sc{iv}, \sc{v} & \sc{ii}, \sc{iii}, \sc{iv}, \sc{v}, \sc{vi} & \sc{ii}, \sc{iii}, \sc{iv}, \sc{v}, \sc{vi}, \sc{vii} & \sc{ii}, \sc{iii}, \sc{iv}, \sc{v}, \sc{vi}, \sc{vii}, \sc{viii} & \sc{iii}, \sc{iv}, \sc{v}, \sc{vi}, \sc{vii}, \sc{viii}, \sc{ix} \\
         		\hline
         	\end{tabular}
         	\label{tab:ionization_stages}
            \begin{minipage}{0.95\linewidth}
                \ignorespaces 
            \end{minipage}
        \end{table*}  

        To calculate synthetic spectra for a given set of stellar and wind parameters, a state-of-the-art stellar atmosphere code that allows deviations from the local thermodynamic equilibrium (LTE) is required. In this work, we employed the PoWR code for non-LTE stellar atmospheres, which has been developed over decades by W.-R. Hamann and co-workers \citep[e.g.,][]{gra1:02,ham1:03,ham1:04,san2:15}. The code has been extensively used to study hot stars ($T\gtrsim\SI{15}{kK}$) at any metallicity \citep[e.g.,][]{san2:14,hai1:14,hai1:15,osk1:11,rei1:14,she1:15}.
        Selected grids of models for both WR and OB stars are already published \citep{ham1:04,san1:12,tod2:15,hai1:19} and are widely used for the analysis of individual stars and populations.
        The setup for this OB model grids follows a similar strategy as detailed in \citet{hai1:19} with few different physical assumptions. For completeness, we repeat all model assumptions.

        Within PoWR, it is assumed that the stellar photosphere and the wind are stationary, spherically symmetric, and in radiative equilibrium. Under these assumptions, the equations of radiative transfer in the CMF and the statistical equilibrium for an expanding non-LTE atmosphere are solved numerically by iteration using the ``approximate lambda operator'' technique. The solution yields the population numbers for each individual atomic level considered within the model atoms.

        The ionization stratification of an element in a model depends on the radiation field, the (electron) temperature, and density. Contrary to the model grids presented by \citet{hai1:19}, the ionization stages included in our grid models are limited to those stages that can be sufficiently populated at the model's effective temperature in order to limit the calculation efforts and boost numerical stability. A complete list of all elements and ionization stages used for a specific temperature range can be found in Table~\ref{tab:ionization_stages}.  For test purposes, we included in a handful of randomly selected models the next higher and lower ionization stage of the iron group elements, with the conclusion that these do not noticeably alter the spectral appearance of a stellar model. 

        For the subsonic regions (i.e., the photosphere) of a stellar atmosphere model, the density stratification and the velocity law are calculated by integrating the hydrostatic equation (hse), which includes the radiation pressure \citep{san2:15} and the pressure by a constant micro-turbulence of $\xi_\mathrm{hse} = \SI{10}{km\,s^{-1}}$. During the calculations in the CMF, in the part of the code that establishes the population numbers, we adopt for the line profile function a Gaussian with a width corresponding to a Doppler velocity of $\varv_\mathrm{dop,\,CMF}=\SI{30}{km\,s^{-1}}$. This should schematically account for thermal and pressure broadening, as well as microturbulence. Such an approximation is usually applied in models for hot-star atmospheres, and has a negligible impact on the resulting spectra \citep[][]{mar1:02,pul1:05,san2:17}.
        
        For the supersonic regions (i.e., the wind), the velocity field is described by a $\beta$ law \citep{cas1:79} of the form
        \begin{equation}
            \varv(r) = \varv_\infty\,\left(1-\dfrac{R_0}{r}\right)^\beta ,
        \end{equation}
        with $R_0\approx R_\ast$ being the connection point between the sub- and supersonic regimes. Note that in the regime where the $\beta$ law is used, $r>R_0$ and thus $\varv(r)>0$. Empirically derived values of $\beta$ range from 0.7 to 1.5, with OB supergiants having preferably higher values \citep[e.g.,][]{kud1:00}. For the grids presented in this paper, we assume an exponent of $\beta=0.8$ as typical for O-type stars \citep{pau1:86,pul1:96}.

        \citet{hil1:91} demonstrated that the winds of hot massive stars are inhomogeneous. In our models, we employ the approximation for optically thin clumps, often called ``microclumping''. The clumping factor $D$ describes the density enhancement within the clumps compared to a smooth wind with an equivalent mass-loss rate \citep{ham1:98}. We presume a depth-dependent clumping, such that in the subsonic regions the wind is smooth, while from the sonic radius onward, the clumping factor increases until it reaches $D=10$ at a radius of $10\,R_\ast$ \citep[e.g.,][]{gra1:05}.

        To generate the synthetic emergent spectrum of a stellar atmosphere model, one needs to integrate the source function in the observer's frame along rays emerging parallel to the observer's line of sight. Within these calculations, the Doppler velocity $\varv_\mathrm{dop}(r)$ consists of the thermal velocity for the individual ion and the ``microturbulence velocity'', denoted as $\xi(r)$. We assume that the microturbulent velocity starts at the photosphere with a value of $\xi_\mathrm{phot}=\SI{10}{km\,s^{-1}}$ and grows in the wind in proportion to the wind velocity $\xi(r) = 0.1\,\varv(r)$. In comparison, \citet{hai1:19} employed the same photospheric value for Galactic and LMC models but adopted $\xi_\mathrm{phot}=\SI{14}{km\,s^{-1}}$ for their SMC grids, introducing minor inconsistencies when models from different grids are used together, for instance in population synthesis. The various types of pressure broadening are accounted for in full detail, as well as the frequency redistribution of photons by electron scattering. The atomic data used in the formal integral are much more detailed than in the CMF calculations, in particular regarding the splitting of multiplets.

        Massive-star winds are permeated by X-rays, which probably originate from embedded shocks due to the instability of the driving mechanism by radiation pressure \citep{cas1:79,fel1:97}. Ionization by X-rays, in particular via the Auger effect, can populate high ionization stages in the wind. From the analysis of O- and B-type stars, it has been shown that the inclusion of X-rays mostly affects the \OVI{} resonance doublet observed in the far-UV spectra of early O-stars and some other wind lines in the UV spectra of B-supergiants \citep[e.g.,][]{osk1:06,bou1:12,ber1:23}. The line strength of photospheric lines is not affected by the inclusion of an X-ray plasma in the wind \citep[e.g.,][]{pau1:01,osk1:11}.
        
    \subsection{Data products}
    \label{sec:products}

        All grid models presented in this paper can be accessed via the user-friendly PoWR web-interface\footnote{\url{www.astro.physik.uni-potsdam.de/PoWR/}} in the ``OB model grids'' category. The new grids presented in this work have the suffix ``-Vd3''. The PoWR website contains various data per selected model in excess to the results presented in this work, including a model's SED, normalized and calibrated synthetic spectra continuously from the UV to infrared with line identifications, as well as synthetic color magnitudes. For the reproducibility of this work and long-term storage, we uploaded the normalized UV and optical spectra used in this paper on Zenodo\footnote{\url{https://doi.org/10.5281/zenodo.18173695}}. For scientific purposes, we strongly advise using the model spectra from the PoWR homepage. Within our model grids, each stellar atmosphere model is named in an ``xx-yy'' format, where ``xx'' is the model's temperature in kilo-Kelvin and ``yy'' is the model's logarithmic surface gravity (in cgs units) multiplied by ten. For example, a model called ``30-44'' has a temperature of $T_\ast=\SI{30}{kK}$ and a surface gravity of $\log(g/(\mathrm{cm\,s^{-2}}))=4.4$.
        
        For each model in our grid, we provide continua as well as calibrated and normalized synthetic spectra covering the range from the UV to the mid-IR (from $\SI{920}{\AA}$ to $\SI{30}{\mu m}$). The resolution of a spectrum is set to 0.3 times the Doppler width of the narrowest line within the spectrum. Since the typical Doppler-velocity in our models is $\varv_\mathrm{dop}\approx\SI{10}{km\,s^{-1}}$, this translates roughly to a resolution of $R\approx\num{100000}$ for our synthetic spectra.        
        
        In addition to the high-resolution synthetic spectra, the total synthetic SED ranging from $\sim\SI{5}{\AA}$ to $\sim\SI{85}{\mu m}$ is provided at low resolution. Furthermore, unreddened photometric magnitudes in the Johnson U, B, and V, Stroemgen $u$, $\varv$, $b$, and $y$, 2MASS J, H, and Ks, and Gaia G$_\mathrm{rp}$, G$_\mathrm{bp}$, and G filters are provided for all models. From our provided data, additional photometry can easily be calculated for any given filter function. Lastly, for each model, the ionizing fluxes for {\sc H}, \HeI{}, \HeII{}, \OII{}, \OIII{}, \NeII{}, and \NeIII{} and Zanstra temperatures for {\sc H} and \HeII{} are provided. 
        
        \begin{figure*}[thb]
            \centering
            \includegraphics[trim={4.5cm 2.cm 5.cm 3.4cm},clip,width=\textwidth]{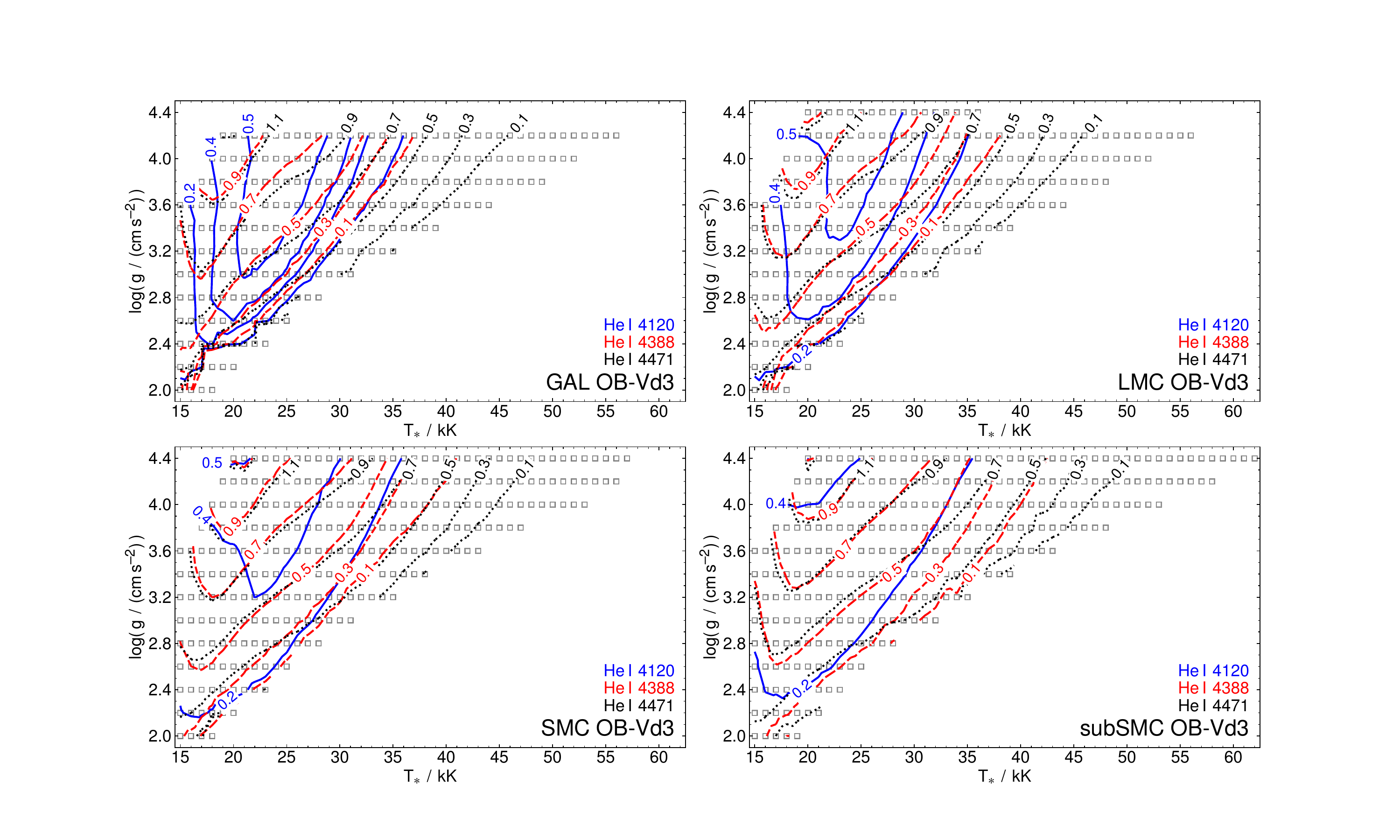}
            \caption{Contour plots depicting the EWs of the \HeI{}\,$\lambda\,4120$, (solid blue) \HeI{}\,$\lambda\,4388$, (dashed red), and \HeI{}\,$\lambda\,4471$ (dotted black) line for all models of the Galactic (GAL; upper left), LMC (upper right), SMC (lower left), and sub-SMC (lower right) OB-Vd3 grid (see Sect.~\ref{sec:products}). Lines of equal EW in $\AA$ are labeled accordingly. Each model within a grid is represented by an open square.}
            \label{fig:HeI4471}
        \end{figure*}
        
    \section{Results}
    \label{sec:results} 
    
        The key temperature diagnostics for hot massive stars are the \HeI{} and \HeII{} lines, along with various metal lines. Typically, line ratios are utilized for spectral classification and first rough temperature estimates. However, these rely on empirical calibrations mainly for Galactic stars. Since the metal lines with their numerous levels can absorb and scatter photons, they are responsible for additional cooling at the surface of a star as well as heating of the inner part of their atmosphere. This effect is known as ``back-warming'' and was first demonstrated by \citet{mih1:78}. The effect has been studied in detail by \citet{rep1:04} using an early version of \textsc{fastwind} \citep{san1:97,her2:02} models of Galactic O stars. \citet{hea1:06} extended the exploration of the back-warming effect in O stars at Galactic and SMC metallicity using large grids of TLUSTY \citep{hub1:95,lan1:03} models. These works showed that back-warming not only affects the line strength of metal lines but also impacts He lines. To further complicate the picture, for a given luminosity, stars with low metallicity tend to be hotter, more compact, and have weaker winds than their higher-metallicity counterparts. Consequently, the resulting different density structure of the stellar atmosphere might lead to different spectral appearances.

    \subsection{EW measurements of temperature diagnostics lines}
    \label{sec:temp}

        For a comprehensive overview of the effect of metallicity on typical diagnostic lines, we computed their EWs. To better compare how the EWs change across the grids, we display them in the form of contour plots as a function of $T$ and $\log g$. Table~\ref{tab:EW} lists the chosen spectral ranges and
        Fig.~\ref{fig:all_lines} displays how the considered lines change as a function of temperature and surface gravity using three models of the LMC grid.
    
    \subsubsection{\HeI{} lines}
    \label{sec:EW_HeI}
    
        \begin{figure*}[thbp]
            \centering
            \includegraphics[trim={4.5cm 2.cm 5.cm 3.4cm},clip,width=\textwidth]{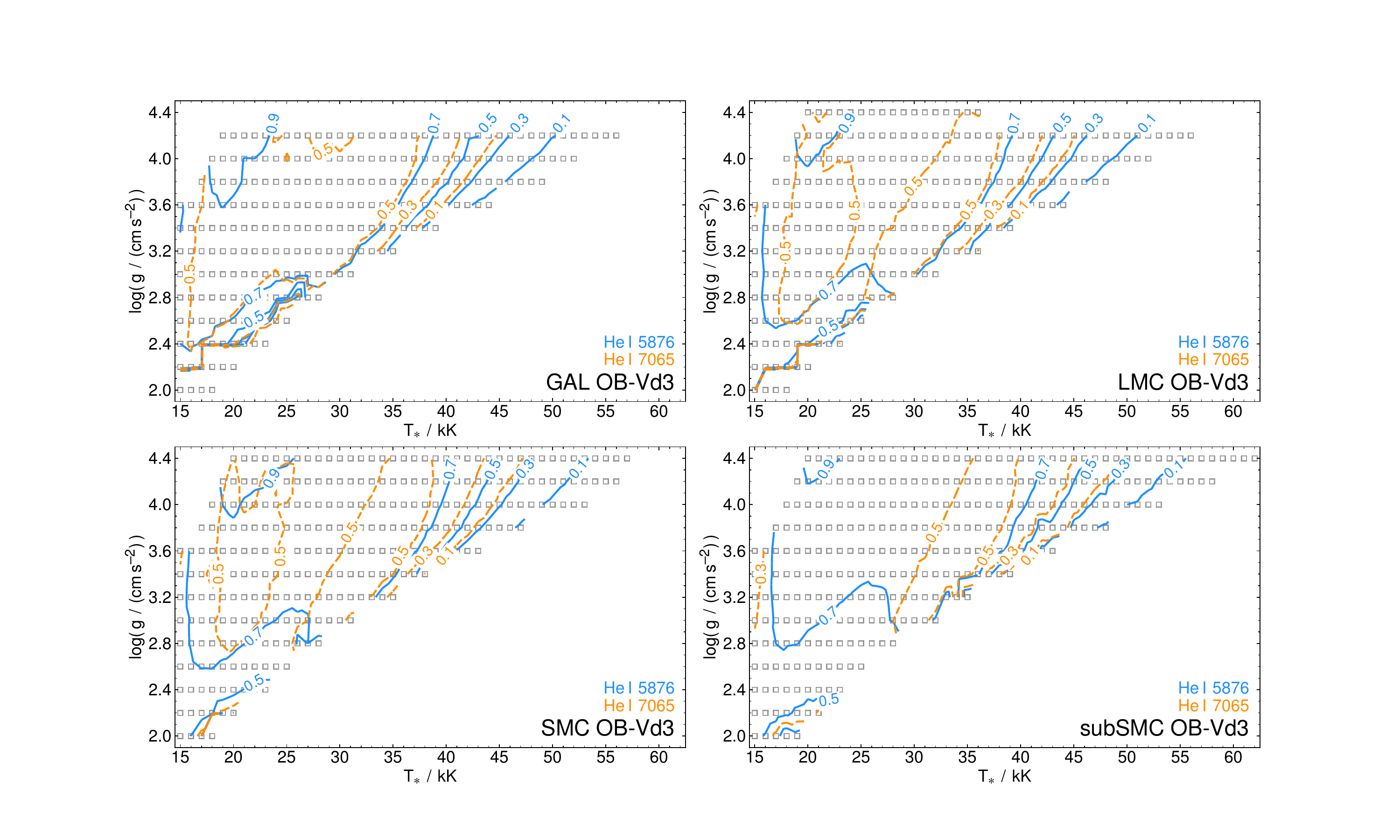}
            \caption{Same as Fig.~\ref{fig:HeI4471} but now for the \HeI{}\,$\lambda\,5876$ (solid blue) and \HeI{}\,$\lambda\,7065$ (dashed orange) lines.}
            \label{fig:HeI5876}
            \centering
            \includegraphics[trim={4.5cm 2.cm 5.cm 3.4cm},clip,width=\textwidth]{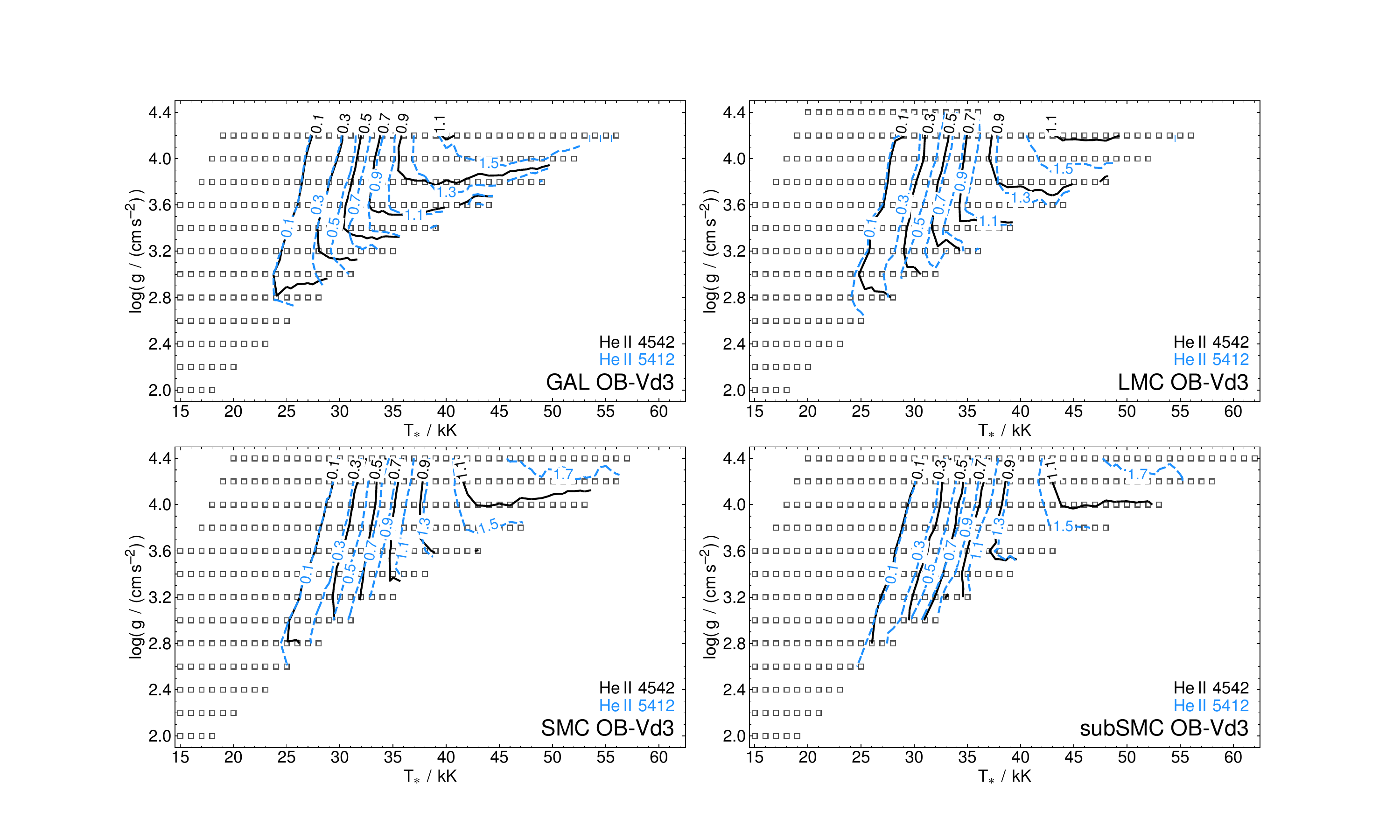}
            \caption{Same as Fig.~\ref{fig:HeI4471} but now for the \HeII{}\,$\lambda\,4542$ (solid black) and \HeII{}\,$\lambda\,5412$ (dashed blue) lines.}
            \label{fig:HeII5412}
        \end{figure*}

        For OB-type stars showing both \HeI{} and \HeII{} lines, their ratio is the ideal indicator for their temperature. However, there are a few limitations: i) some of the He-lines are blended with other lines, such as the Balmer lines, and ii) \HeI{} singlet lines with transitions involving the 1s2p $^1$P level, exhibit discrepancies to observations within a specific temperature range for unknown reasons \citep[see ][]{naj1:06}. Consequently, these lines are not considered here. 

        Figure~\ref{fig:HeI4471} depicts how the EWs of key diagnostic \HeI{} varies across the $T-\log g$ plane. The \HeI{}\,$\lambda\,4471$ is one of the most prominent lines in OB-stars spectra, which is also reflected in the broad temperature range it covers. However, one can also see a strong correlation of the EW of \HeI{}\,$\lambda\,4471$ with surface gravity, with the strongest lines expected in the spectra of B dwarfs around $T\approx\SI{20}{kK}$.\citet{hea1:06} reported that in their TLUSTY models at Galactic metallicity, due to the effect of back-warming, the \HeI{}\,$\lambda\,4471$ is strengthened in models with $T\leq\SI{40}{kK}$ and weakened in models with $T\geq\SI{40}{kK}$ when compared to their SMC and pure H-He models. They report that a SMC model with the same EW as a Galactic model is about $\SIrange{1}{3}{kK}$ hotter. A similar shift is observed in our models when comparing the GAL and SMC grids. In our GAL and LMC grid, one can see discontinuities in the contour plots at $\log g \lesssim2.6$. These models have such strong mass-loss rates that they lack \HeI{}\,$\lambda\,4471$ or show it in emission, underlining the effect of mass-loss on temperature diagnostic lines. 

        \citet{sot1:11} suggested using the \HeI{}\,$\lambda\,4388$ line as an additional diagnostic for the spectral classification of late O-type stars. As shown in  Figure~\ref{fig:HeI4471}, for temperatures below $T\lesssim\SI{30}{kK}$ the EW of \HeI{}\,$\lambda\,4388$ increases in a similar manner as that of \HeI{}\,$\lambda\,4471$. However, at higher temperatures, the \HeI{}\,$\lambda\,4388$ line rapidly weakens and eventually disappears, highlighting its strong temperature sensitivity in the late O-stars regime.

        For early B-type stars the \HeI{}\,$\lambda\,4120$ line is commonly employed alongside metal lines for spectral classification \citep[][]{gra2:09,neg1:24}. From Fig.\,\ref{fig:HeI4471}, one can see that this line reaches the maximum EW at $T\approx\SI{25}{kK}$ at solar metallicity and shows a weaker dependence on $\log g$ than the other two \HeI{} lines presented above. At lower metallicity, the \HeI{}\,$\lambda\,4120$ line is present in the models at lower temperatures, and interestingly, the pronounced peak in EW around $T\approx\SI{25}{kK}$ becomes less distinct. This line seems very sensitive to the effect of metallicity in the form of back-warming, as at sub-SMC metallicity, the line is a factor of two weaker than in our GAL grid.

        Another \HeI{} line used for the spectral analysis of massive stars is \HeI{}\,$\lambda\,5876$. The variation in EW across the grid models is shown in Fig.~\ref{fig:HeI5876}. Surprisingly, for $T\lesssim \SI{37}{kK}$ and $\log(g/(\mathrm{cm\,s^{-2}}))\gtrsim3.0$ the EW of this line remains almost constant, making it a weak diagnostic line for B-stars. However, for $T\approx\SIrange{37}{50}{kK}$ the line is sensitive to $T$ and $\log g$. In particular, in the sub-SMC grid, where the effect of back-warming is the weakest, this line should be detectable in the spectra of main-sequence stars as hot as $T\approx\SI{55}{kK}$. This makes the \HeI{}\,$\lambda\,5876$ a valuable temperature diagnostic for hot low-metallicity stars.

        The \HeI{} line at $\lambda\,7065$ can be an additional constraint for the temperature estimate. As shown in Fig.~\ref{fig:HeI5876}, this line appears relatively insensitive to $T$ and $\log g$ at solar metallicity. However, at LMC and SMC metallicity, the EW has two maxima around $T\approx\SI{20}{kK}$ and $T\approx\SI{32}{kK}$. At sub-SMC metallicity, only one maximum in EW at $T\approx\SI{32}{kK}$ remains. 

        Our models demonstrate that \HeI{} lines have different sensitivities to $T$ and $\log g$. In the case of an absence of metal lines, such as in our sub-SMC grid (see Sect.~\ref{sec:metal}), one can use a combination of different \HeI{} lines to roughly constrain the stellar temperature.

    \subsubsection{\HeII{} lines}

        Figure~\ref{fig:HeII5412} illustrates how the EW of the \HeII{}\,$\lambda\,4541$ line varies across the grid models. The EW of this line shows only a weak dependence on 
        $\log g$ and increases quickly with temperature, making it an excellent diagnostic line for stars with temperatures between $\SIrange{25}{45}{kK}$.  Only at temperatures above $\SI{40}{kK}$, the \HeII{} lines become more sensitive to $\log g$ and almost insensitive to $T$. 
        
        \citet{hea1:06} report that in their TLUSTY model grids, the \HeII{}\,$\lambda\,4541$ is strengthened with increasing metallicity due to the effect of back-warming and that this increase in EW is strongest at lower temperatures. While our models also show this trend, they also show that with increasing metallicity (i.e., more efficient back-warming), the sensitivity of this \HeII{} line to $\log g$ increases, most notably at $T\lesssim\SI{38}{kK}$.
        
        The \HeII{}\,$\lambda\,5412$ as well as other isolated \HeII{} lines in our grid exhibit similar sensitivity to $T$ and $\log g$ as the \HeII{}\,$\lambda\,4541$ line. However, the lines have different intensities and responses to different combinations of $T$ and $\log g$. For instance, in Fig.~\ref{fig:HeII5412} one can see that the \HeII{}\,$\lambda\,5412$ line has much larger EWs and extends to much higher $T$ when compared to the \HeII{}\,$\lambda\,4541$ line.
        
        \begin{figure*}[thbp]
            \centering
            \includegraphics[trim={4.5cm 2.cm 5.cm 3.4cm},clip,width=\textwidth]{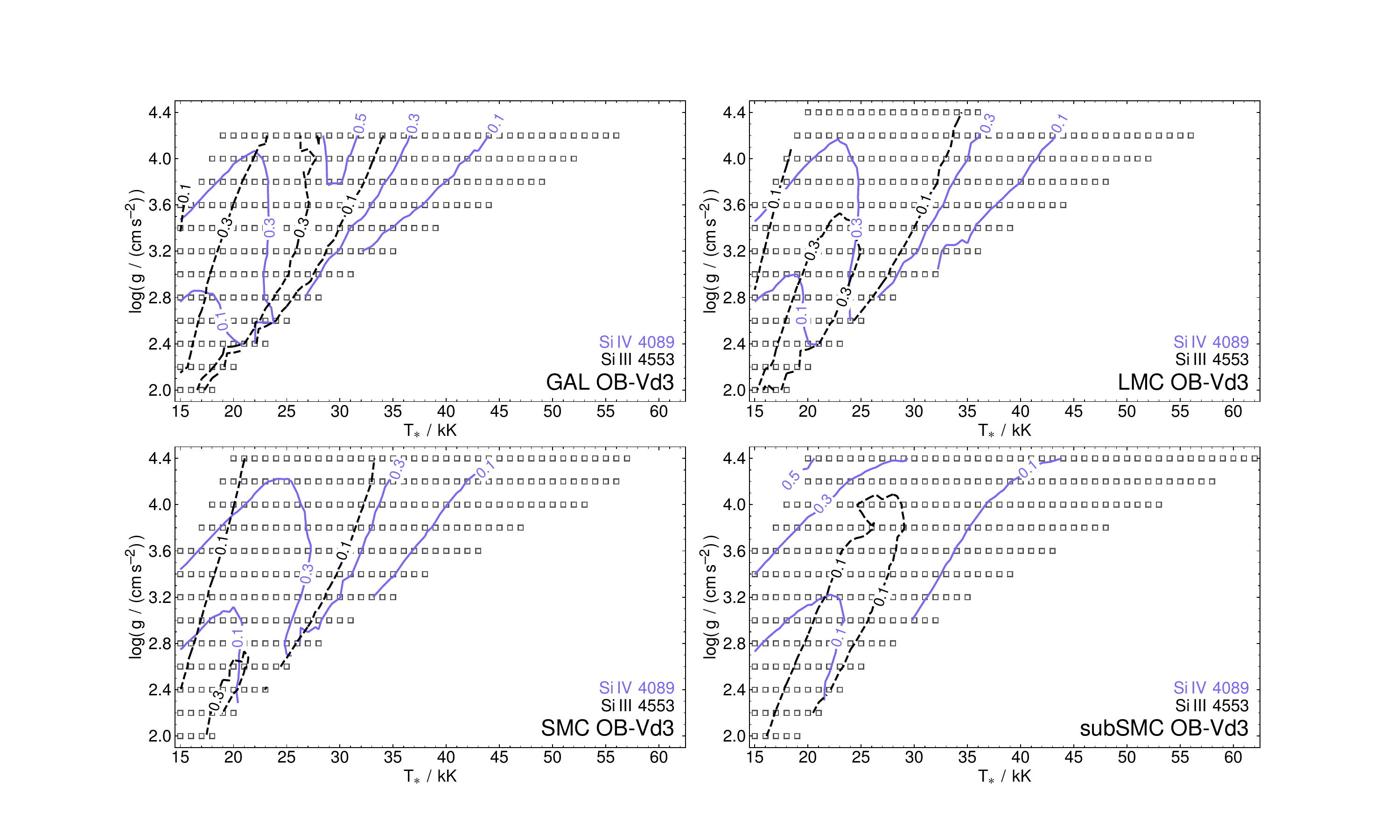}
            \caption{Same as Fig.~\ref{fig:HeI4471} but now for the \SiIII{}\,$\lambda\,4553$ (dashed black) and \SiIV{}\,$\lambda\,4089$ (solid purple) lines.}
            \label{fig:SiIII4553}
            \vspace{1ex}
            \centering
            \includegraphics[trim={4.5cm 2.cm 5.cm 3.4cm},clip,width=\textwidth]{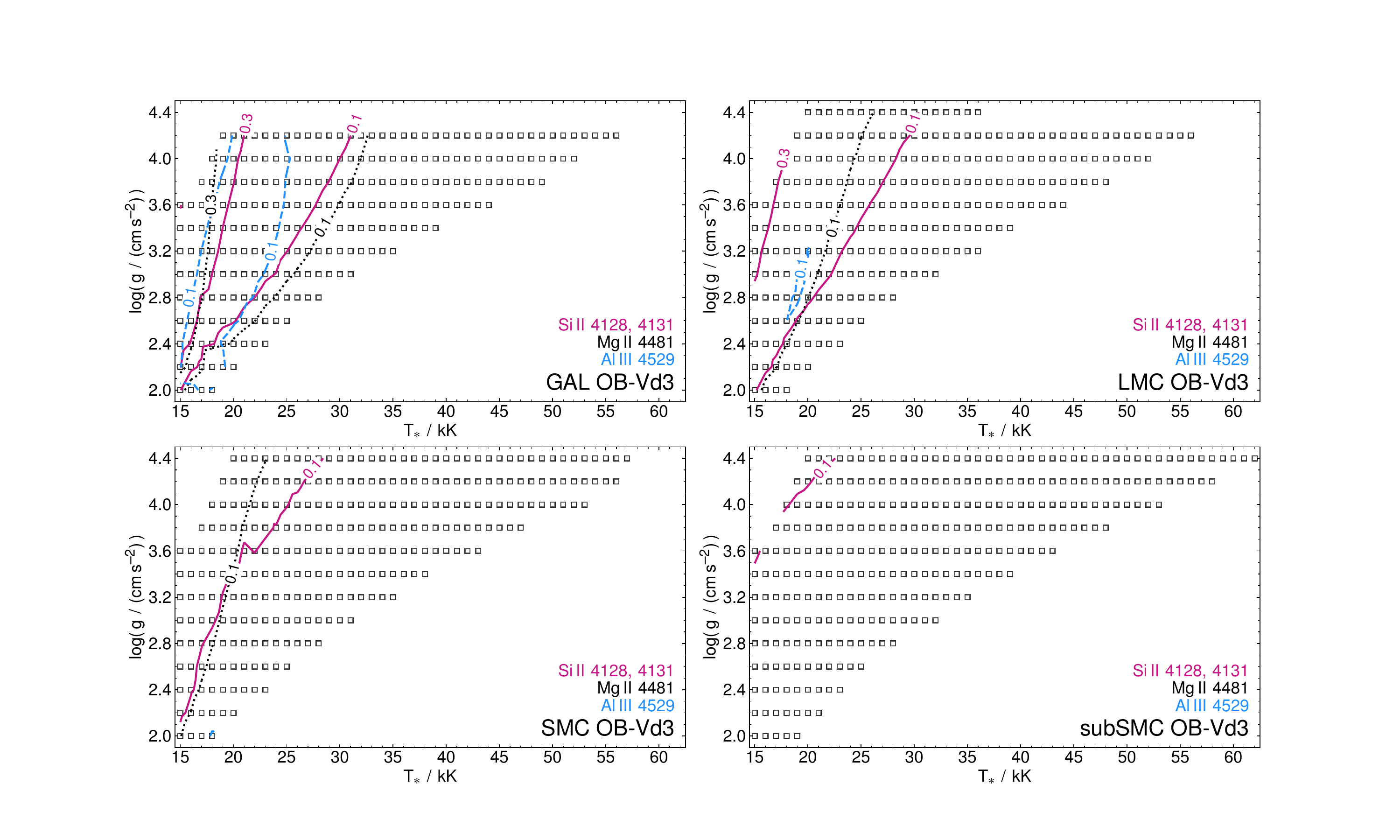}
            \caption{Same as Fig.~\ref{fig:HeI4471} but now for the \SiII{}\,$\lambda\lambda\,4128,4131$ (solid magenta), \MgII{}\,$\lambda\,4481$ (dotted black), and \AlIII{}\,$\lambda4529$ (dashed blue) lines.}
            \label{fig:MgII4481}
        \end{figure*}

    \subsubsection{Metal lines}
    \label{sec:metal}

        \HeI{} and \HeII{} lines are crucial to constrain the temperatures of OB-type stars. However, as demonstrated above, \HeII{} lines disappear in the spectra of stars with $T\lesssim\SI{30}{kK}$ (i.e., B-type stars). Commonly, for such cool stars, metal lines of silicon, magnesium, and aluminum in combination with the \HeI{} lines are employed as temperature constraints \citep[e.g.,][]{sot1:11,eva1:15,mce1:15,neg1:24}. To comprehend the capabilities and limitations of these lines across varying temperatures and metallicities, we also calculated the EWs for selected, unblended Si and Mg lines.
        
        In Fig.~\ref{fig:SiIII4553}, the change in EW of \SiIII{}\,$\lambda4553$ and \SiIV{}\,$\lambda4089$, some of the strongest metal lines in OB-type stars, is illustrated. These \SiIII{} and \SiIV{} lines are weakly sensitive to $\log g$, and peak at $T\approx\SI{20}{kK}$ and $T\approx\SI{30}{kK}$, respectively. As metallicity decreases, the \SiIII{} and \SiIV{} lines get weaker but remain detectable in several of our sub-SMC models. This makes these silicon lines important additional temperature diagnostics of late-O and B-type stars, even at extremely low metallicities. However, in rotating stars, these diagnostics might disappear in the noise of the observation, preventing an accurate temperature determination.    
        
        The EWs of \MgII{}\,$\lambda\,4481$, \SiII{}\,$\lambda\lambda\,4128,4131$, and \AlIII{}\,$\lambda4529$ across the $T-\log g$ plane are shown in Fig.~\ref{fig:MgII4481}. 
        Most of these lines are only present in the models with the lowest temperatures covered by our grid, making them particularly interesting for the analysis of B-type stars.
        Their strength rapidly decreases with metallicity: the \AlIII\ line is no longer present below LMC metallicity while the \MgII{} and \SiII{} lines vanish below SMC metallicity. \citet{gul1:22} presented optical spectra of a small sample of OB stars in Sextrans A ($Z\sim1/20\,\zsun$) and reported measured EW of \MgII{}\,$\lambda\,4481$ and \SiII{}\,$\lambda\,4128$ between $\SIrange{0.2}{0.8}{\AA}$. However, a visual inspection of their spectra reveals no clear signature of these lines, being in alignment with our model predictions.
        
        For the spectral classification of hot stars where \HeI{} lines are absent, \citet{wal1:04} suggested employing N lines from different ionization stages. However, theoretical modeling efforts by \citet{gon1:12} demonstrated that these lines are sensitive to surface gravity, metallicity, mass-loss, nitrogen abundance, and the treatment of dielectronic recombination.

        \subsection{EW ratios used for spectral type classification}
        \begin{figure*}[thb]
            \centering
            \includegraphics[trim={4.2cm 2.cm 4.8cm 1.7cm},clip,width=\textwidth]{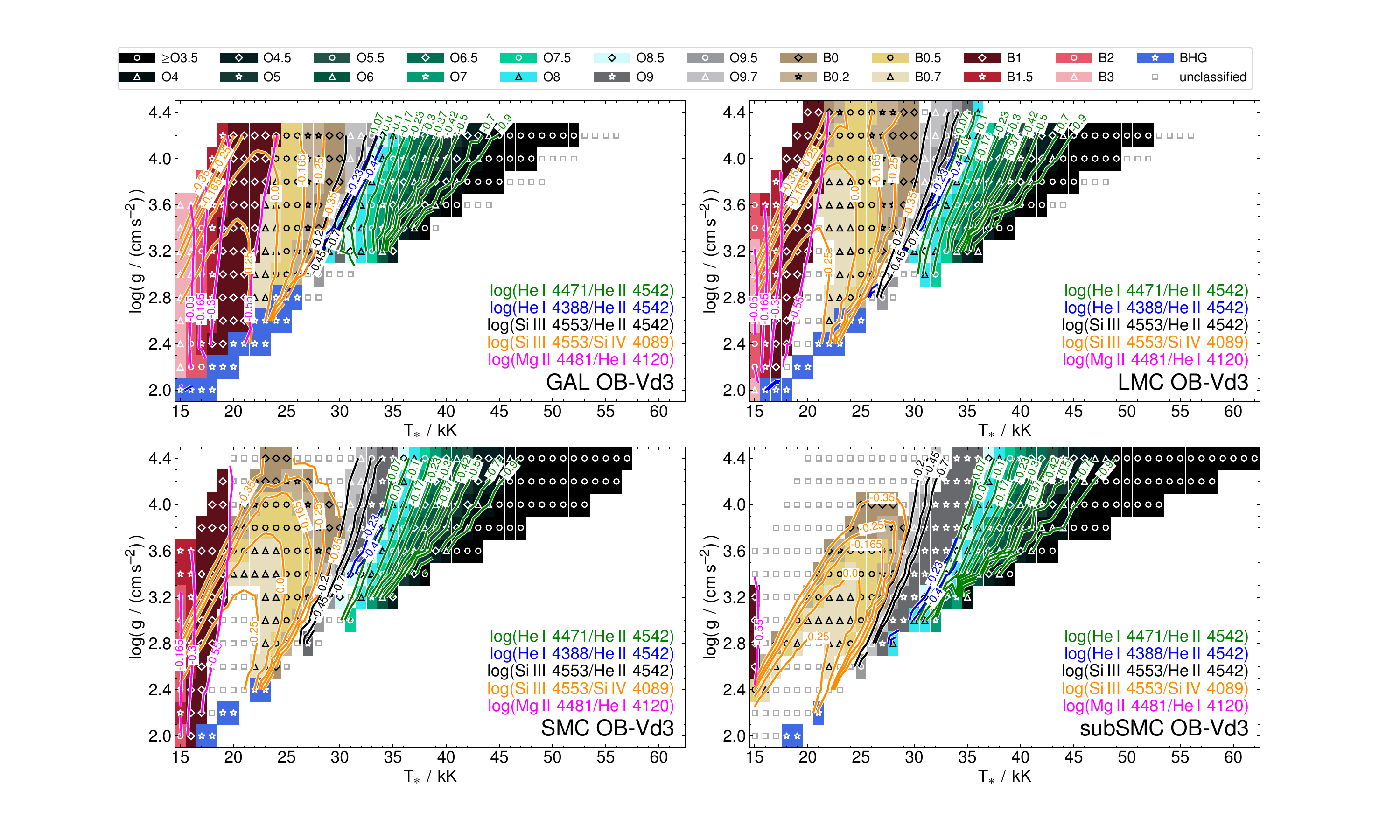}
            \caption{Contourplots showing the change in the EW ratios using the observationally established classification boundaries (see Table~\ref{tab:ratios} for all models of the GAL (upper left), LMC (upper right), SMC (lower left), and subSMC (lower right) OB-Vd3 grid (see Sect.~\ref{sec:grid}. Shown are the logarithmic EW ratios of $\log(\HeI{}\,\lambda4471/\HeII{}\,\lambda4542)$ (green lines), $\log(\HeI{}\,\lambda4388/\HeII{}\,\lambda4542)$ (blue lines), $\log(\SiIII{}\,\lambda4553/\HeII{}\,\lambda4542)$ (black lines), $\log(\SiIII{}\,\lambda4553/\SiIV{}\,\lambda4089)$ (orange lines), and $\log(\MgII{}\,\lambda4481/\HeI{}\,\lambda4120)$ (pink lines). In the background, colored boxes indicate the spectral type of a model where possible. If none of the classification criteria were applicable, the model is unclassified and shown as an open gray square.}
            \label{fig:line_ratios}
        \end{figure*}
    
        For the purpose of spectral classification, and thus a first-order estimate of the temperature, line ratios in the optical are commonly employed. Note that different approaches exist in the literature: comparisons with stellar template spectra \citep[e.g.,][]{fit1:91,cas1:08,mai1:19}, line intensity ratios \citep[e.g.,][]{wal1:90,sot1:11,bes1:25}, and EW ratios \citep[e.g.,][]{mat1:88,mar2:18}. Because of technical limitations in the past, the optical spectral classification is typically based on the rather narrow wavelength range from $\approx\SIrange{4000}{5000}{\AA}$.
        
        For the classification of O-type stars, \citet{mat1:88} suggested to use the $\log(\HeI{}\,\lambda4471/\HeII{}\,\lambda4542)$ EW ratio. However, \citet{mar2:18} has re-investigated these limits in combination with the line ratios for O-type stars later than $>$O8 as suggested by \citet{sot1:11} and found that various diagnostics exhibit substantial scatter. As a result, the observed line ratios often overlap with previously defined classification boundaries and cannot be used. For B-type stars \citet{did1:82} noted that the EWs of a given line depend not only on $T$, but also on the luminosity class (and thus $\log g$). Consequently, for the spectral classification of B stars, different lines are used depending on the luminosity class of the targeted star \citep[e.g.,][]{neg1:24}.

        To mitigate these limitations and establish a quantitative spectral classification framework directly comparable to our theoretical models, we compiled a homogeneous set of observational templates. Specifically, we adopted Galactic O-star templates from \citet{mai1:11, mai1:13, mai1:16} and Galactic B-star templates from \citet{neg1:24}. Note that the O-star templates already consider the updated classification for late O-types from \citet{sot1:11} and \citet{sot1:14}. Both, the O- and B-star samples were cleaned of known binaries and used to derive EW-based line ratios. Based on the classification schemes of \citet{mat1:88}, \citet{sot1:11}, and \citet{neg1:24}, we employed the $\log(\HeI{}\,\lambda4471/\HeII{}\,\lambda4542)$, $\log(\HeI{}\,\lambda4388/\HeII{}\,\lambda4542)$, $\log(\SiIII{}\,\lambda4553/\HeII{}\,\lambda4542)$, and $\log(\SiIII{}\,\lambda4553/\SiIV{}\,\lambda4089)$ for O and early-B type stars. For B1 to B3 stars, we tested several line ratios and discovered that the $\log(\MgII{}\,\lambda4481/\HeI{}\,\lambda4120)$ ratio is largely insensitive to the luminosity class (and thus $\log g$). A summary of the adopted line-ratio criteria for each spectral type is provided in Table~\ref{tab:ratios}, and their correspondence with observational data is illustrated in Fig.~\ref{fig:obs_ratios}. We note that the derived EW ratios can be sensitive to the adopted integration limits (see Table~\ref{tab:EW}). This is particularly relevant for \HeI{}\,$\lambda4471$, which lies close to \MgII{}\,$\lambda4481$. Increasing the integration window around \HeI{}\,$\lambda4471$ results in systematically larger measured EWs and therefore larger line ratios, partly due to the inclusion of flux from the nearby \MgII{} feature. As the integration limits used in previous studies are often not specified, differences in the adopted measurement procedure can lead to systematic offsets between published EW-ratio calibrations. Consequently, our empirically derived EW-ratio criteria are not expected to coincide exactly with those reported in earlier works.

    \subsubsection{Solar metallicity}

        Figure~\ref{fig:line_ratios} shows the EW ratios introduced above and the resulting spectral classifications for each model across our different model grids. For the GAL model grid, the adopted classification boundaries generally yield consistent spectral types, as most EW ratios overlap in well-defined regions without leading to contradictory classifications. One exception is the $\log(\HeI{}\,\lambda4388/\HeII{}\,\lambda4542)$ ratio. For models with high $\log g$, this ratio overlaps with the classification boundaries of the same spectral type as defined by $\log(\HeI{}\,\lambda4471/\HeII{}\,\lambda4542)$, while at lower $\log g$ it coincides with the later O9 classification regions based on $\log(\SiIII{}\,\lambda4553/\HeII{}\,\lambda4542)$. At the highest temperatures covered by the grid, several O-type models cannot be classified due to the absence of detectable \HeI{} lines. As noted by \citet{wal1:04} for Galactic stars, nitrogen lines could serve as alternative diagnostics in this regime.
        
        For B0 to B1 stars the $\log(\SiIII{}\,\lambda4553/\SiIV{}\,\lambda4089)$ ratio shows strong temperature sensitivity from $\SIrange{22}{30}{kK}$. Interestingly, this ratio remains measurable in our models at lower temperatures and high surface gravity. However as demonstrated in Fig.~\ref{fig:SiIII4553}, the \SiIV{}\,$\lambda4089$ disappears in B giants and supergiants. Consequently the $\log(\MgII{}\,\lambda4481/\HeI{}\,\lambda4120)$ ratio, which is mostly insensitive to surface gravity, is instead used for the classification of B1–B3 stars.
        
        Finally, models with the lowest $\log g$ at a given $T$ lie beyond the Humphreys–Davidson limit \citep{hum1:79} (see Fig.~\ref{fig:HRD}). According to the employed mass-loss recipe, these stars have strong stellar winds that significantly impact the ionization structure of their atmospheres, leading to either an Of/WN, OB Ia+, or late WN classification when \Hbeta{} and other spectral lines appears in emission (EW $< 0\,\AA$). For simplicity, we labeled these stars in our grid as blue hypergiants (BHG).

    \subsubsection{LMC metallicity}
    
        For the LMC grid, shown in the upper right panel of Fig.\,\ref{fig:line_ratios}, the spectral classification still works reasonably well. As discussed in Sect.~\ref{sec:results}, the reduced line blanketing via back-warming, and the higher density structure of stellar atmospheres at lower metallicity shifts the same EW of helium lines to higher $T$. Consequently, EW ratios based solely on helium lines are shifted toward higher $T$. In contrast, metal lines weaken with decreasing metallicity due to the lower abundances, resulting in EW ratios involving metal lines being shifted toward lower $T$. This combined effect already introduces classification ambiguities at LMC metallicity. In particular, models with low $\log g$ and $T \approx \SI{19}{kK}$ fall outside the established classification boundaries and can no longer be reliably classified. This limitation was already noted in a study of B supergiants in the LMC by \citet{fit1:91}, who suggested revising the classification criteria and direct comparisons with spectral templates in such cases.

    \subsubsection{SMC metallicity}
    
        At SMC metallicity (lower panel of Fig.~\ref{fig:line_ratios}) the situation becomes more pronounced. For O-type stars, the reduced back-warming shifts EW ratios based solely on helium lines toward higher $T$ by $\sim$$\SI{3}{kK}$. This is consistent with shifts reported in both theoretical and empirical studies of O stars in the SMC \citep[][]{bou1:03,eva1:04,mas1:04,mas1:05,hea1:06,mok1:06,mas1:09}. Extensive analyses by \citet{mas1:04,mas1:05,mas1:09} have also provided detailed spectral classifications for the SMC's O-type stars. For the earliest O-type stars, these authors employed the morphology of nitrogen lines as proposed by \citet{wal1:04}. However, this approach becomes increasingly uncertain at low metallicity due to the scarcity of very early-type O stars in the SMC and the sensitivity of nitrogen line strengths to both abundance variations (reflecting prior evolutionary processes) and atmospheric conditions. Our models instead predict that \HeI{} lines should remain detectable even at temperatures as high as $T=\SI{55}{kK}$, potentially enabling a more robust spectral classification of the earliest O-type stars at low metallicity without relying on lines affected by CNO processing.

       For late-O and early-B stars at SMC metallicity, commonly used metal-line ratios begin to fail, particularly for stars at both high and low $\log g$ (i.e., main-sequence and supergiants). Additionally, the temperature range associated with spectral types B0 to B0.7 broadens significantly. \citet{len1:97} investigated B supergiants in the SMC and proposed alternative classification criteria by calculating the EWs of He, C, N, O, Mg, and Si lines and identifying correlations that improve temperature estimates. Notably, they found trends similar to those seen in our models: \MgII{} becomes prominent only in stars later than B1, while \SiIII{} can already be detected at spectral type B0. Based on their results and supported by our theoretical EW measurements (see Fig.~\ref{fig:SiIII4553}), we conclude that, to mitigate classification gaps at low metallicity, one should preferentially use EW diagnostics that do not rely on the simultaneous presence of multiple ionization stages of the same element within the observable spectral range.

    \subsubsection{sub-SMC metallicity}
    
       For our subSMC model grid, presented in the lower right of Fig.~\ref{fig:line_ratios}, the O-type stars classified using helium-line ratios follow trends very similar to those seen in the SMC models, namely that the hottest stars in the model grid can be classified as O3.5 to O4 stars due to the presence of the \HeI{} lines. This is in agreement with observations of massive stars in Sextans A, where the spectra presented by \citet{lor1:22} clearly display the presence of \HeI{}$\lambda4471$ even for O4\,V stars.
       
       Another emerging feature in this model grid is that the temperature range associated with O9 stars becomes noticeably broader. This behavior arises from the transition between diagnostics based purely on helium lines and those involving silicon lines, which become significantly weaker at these low abundances. For B-type stars, the spectral classification of the B0 to B0.7 covers a wide temperature range due to the overall lack of the \MgII{} line. Nevertheless, as in the SMC grid, at least one of the silicon lines (\SiIII{} or \SiIV{}) is typically still detectable in many models (see Fig.~\ref{fig:SiIII4553}). Considering their EWs individually may therefore provide a partial means of classification. However, to obtain robust temperature estimates at such low metallicities, detailed stellar atmosphere modeling becomes essential. This approach must account for the differing sensitivities of both metal and \HeI{} lines to changes in stellar parameters, as already mentioned in Sect.~\ref{sec:temp}.
                
        We emphasize that the spectral classification presented above is based on nonrotating stellar atmosphere models under idealized conditions, assuming no noise contamination. In practice, however, rotational broadening, which is likely more pronounced at lower metallicity, and the limited signal-to-noise ratios of observations, particularly for stars in distant galaxies, can hinder the detection of weak spectral lines used for classification.
        
        \begin{figure*}[thbp]
            \centering
            \includegraphics[trim={3.5cm 0cm 3.cm 0cm},clip,width=0.9\textwidth]{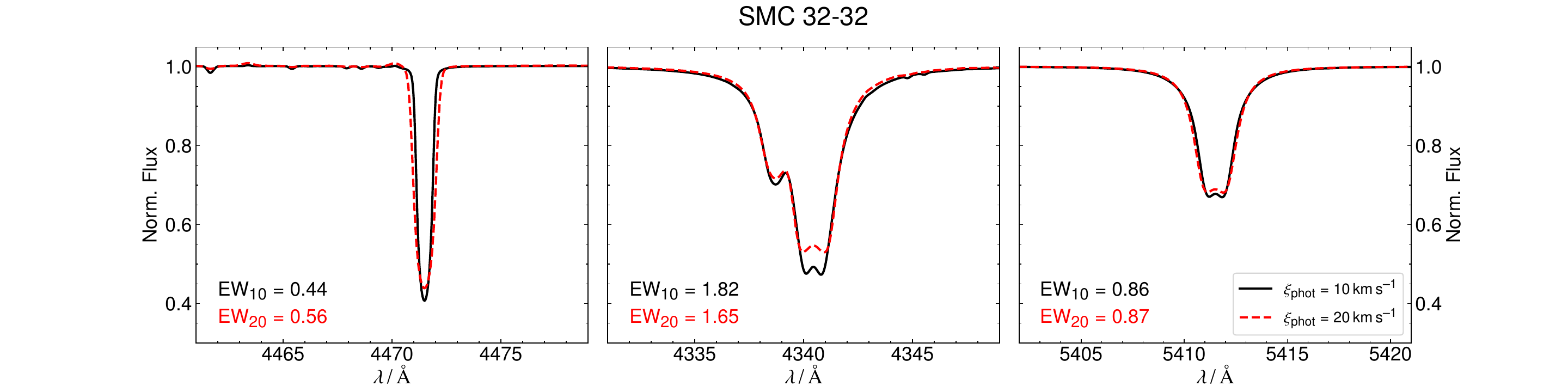}
            \includegraphics[trim={3.5cm 0.2cm 3.cm 0cm},clip,width=0.9\textwidth]{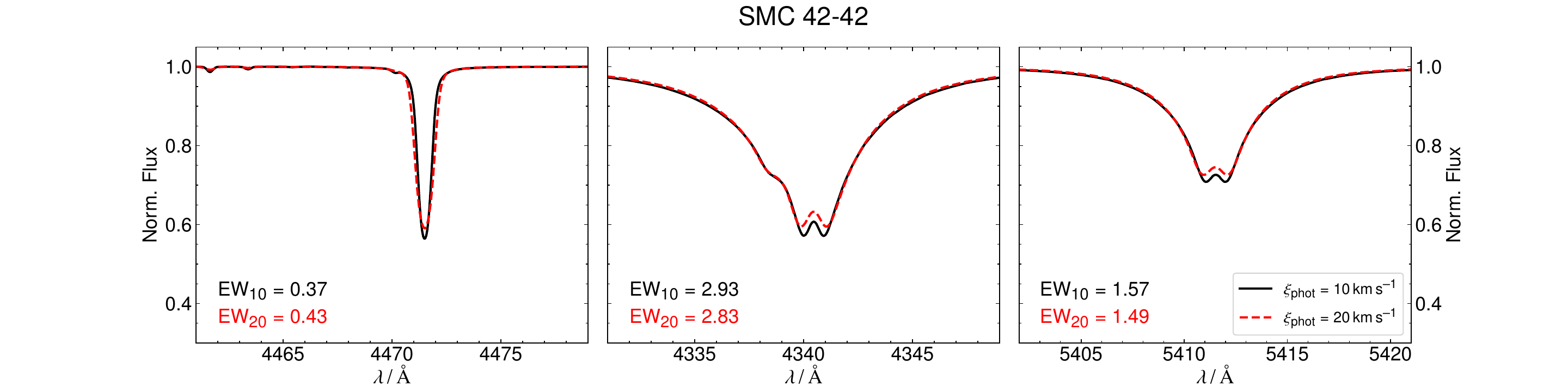}
            \caption{Comparison of the impact of different microturbulent velocities on selected lines of the SMC 32-32 (upper panels) and SMC 42-42 model (lower panels). The model from our stellar atmosphere grid ($\xi_\mathrm{phot}=\SI{10}{km\,s^{-1}}$) is shown as solid black line and the same model with increased microturbulent velocity ($\xi_\mathrm{phot}=\SI{20}{km\,s^{-1}}$) is shown as dashed red line.}
            \label{fig:vmic20}
            \vspace{2ex}
            \centering
            \includegraphics[trim={4.2cm 2.cm 4.8cm 1.7cm},clip,width=\textwidth]{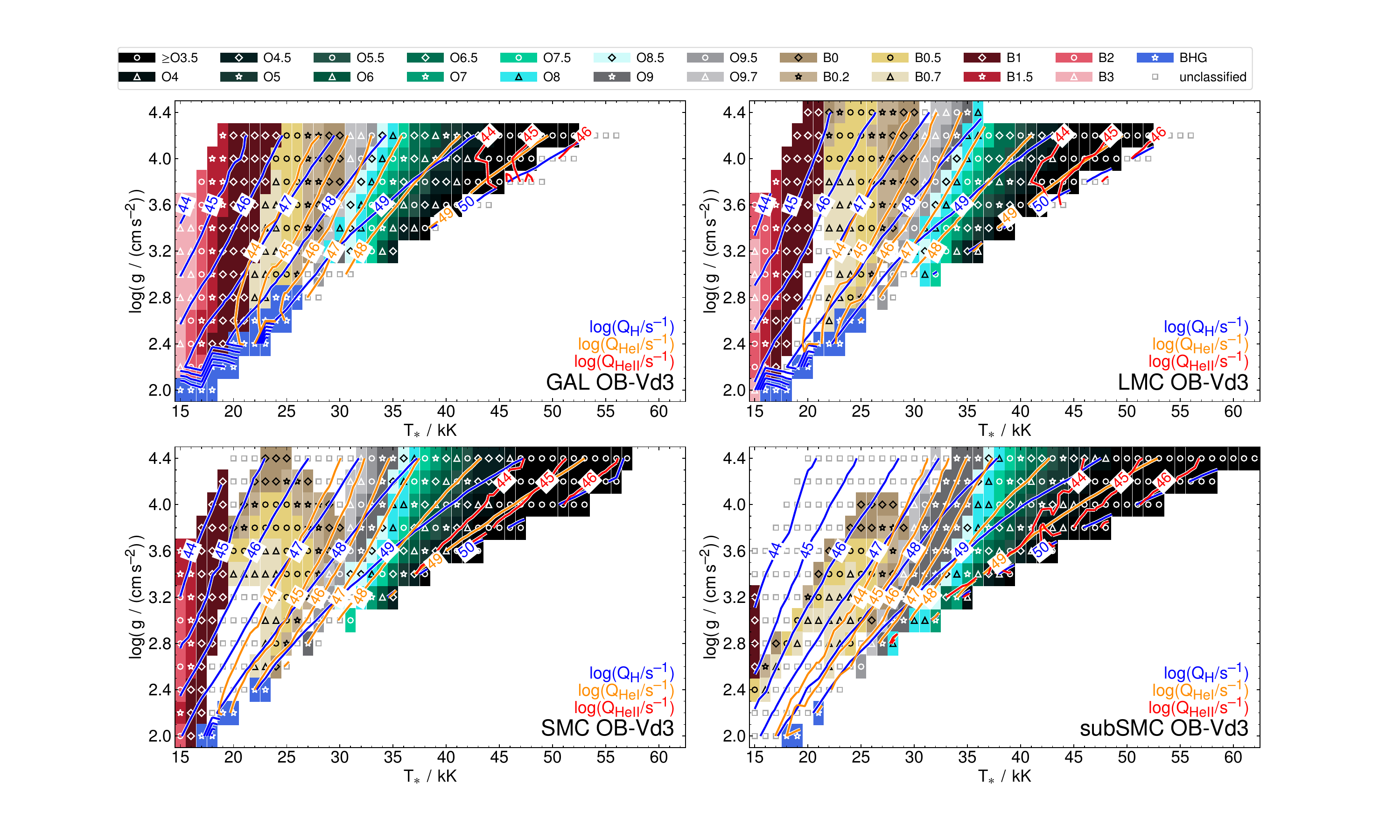}
            \caption{Contourplots showing the change in the hydrogen- and helium ionizing fluxes for all models of the GAL (upper left), LMC (upper right), SMC (lower left), and subSMC (lower right) OB-Vd3 grid. Shown are the logarithmic ionizing fluxes of {\sc H} (blue lines), \HeI{} (orange lines), and \HeII{} (red lines). In the background, colored boxes indicate the spectral type of a model where possible. If none of the classification criteria were applicable, the model is unclassified and shown as an open gray square. }
            \label{fig:logQ}
        \end{figure*}
    
    \subsection{Impact of microturbulence on diagnostic lines}
    \label{sec:discuss}

        The structure of the quasi-hydrostatic layers has a large impact on the emergent absorption-line spectrum. One important factor impacting the quasi-hydrostatic layers is the back-warming effect due to all relevant opacities, in particular from the iron group elements \citep[e.g.,][]{mih1:78}. These opacities are also essential for evaluating the radiation pressure, which PoWR takes into account in all detail for the quasi-hydrostatic equation. 
        
        A further term in that equation is the pressure exerted by the turbulent motion. This pressure increases the pressure scale height, and thus effectively acts like decreasing the surface gravity ($\log g$) \citep[see also][]{mce1:98,san2:15,san2:17}. For the model grids presented here, we assume a microturbulent velocity of ${\xi_\mathrm{hse} = \SI{10}{km\,s^{-1}}}$, which is constant throughout the photosphere. 
        
        The same photospheric microturbulence is taken into account for calculating the emergent spectra ($\xi_\mathrm{phot}=\SI{10}{km\,s^{-1}}$). For the wind, which is not relevant for the photospheric absorption lines discussed in this paper, we assume that the microturbulence grows as 10\% of the local wind velocity (for more details, see Sect.~\ref{sec:PoWR}). 
        
        Our choice for the photospheric microturbulence of $\SI{10}{km\,s^{-1}}$ is schematic; individual stars might show different values. From fitting sharp metal lines in high-resolution observations, it is known that the microburbulence is in the range $\xi=\SIrange{5}{30}{km\,s^{-1}}$\citep{har1:70,lam1:72,duf1:72,len1:91,gie1:92}. 
        
        Naturally, higher microturbulence slightly broadens the Doppler cores of the absorption lines. To test the impact of microturbulence on the emergent spectra, we picked up a handful of models from our grids and recalculated them with $\xi_\mathrm{hse}=\xi_\mathrm{phot}=\SI{20}{km\,s^{-1}}$. Two examples are illustrated in Fig.~\ref{fig:vmic20}. There are only small differences in the EWs of the hydrogen and helium lines. These differences mostly correspond to shifting the contour plots along the $\log\,g$ axis, as expected. 

    \subsection{Ionizing fluxes}

        The ionizing output of a star is primarily determined by its temperature and luminosity. However, in stars with strong stellar winds, the wind can become (partially) optically thick for \HeII{} and in extreme cases even for \HeI{} and {\sc H} ionizing photons. As a result, a significant fraction of ionizing photons is absorbed within the wind itself, where it contributes to driving the outflow rather than escaping to ionize the surrounding medium. To quantify the dependence of the hydrogen- ($\log Q_\mathrm{H}$) and helium-ionizing photon fluxes ($\log Q_\ion{He}{I}$ and $\log Q_\ion{He}{II}$) with $T$ and $\log g$, and spectral type, we computed contour plots, displayed in Fig.~\ref{fig:logQ}.

        An inspection of Fig.~\ref{fig:logQ} shows that, as expected, all models in our grid emit more than $10^{44}$ hydrogen-ionizing photons per second. In the GAL model grid, only stars earlier than $\gtrsim$O6 reach $\log(Q_\mathrm{H}/\mathrm{s^{-1}})\gtrsim49$, and are thus capable of significantly ionizing hydrogen in their immediate surroundings. The precise spectral-type threshold depends on the evolutionary stage represented by a given stellar atmosphere model. In particular, models with lower $\log g$ correspond to more evolved, highly luminous supergiants (see Fig.~\ref{fig:HRD}). Note that once a star develops a dense, optically thick stellar wind, as in the case of BHG stars, a substantial fraction even of the {\sc H} ionizing radiation is absorbed within the wind and contributes to driving it, rather than escaping to ionize the surrounding medium. At lower metallicities, such as in the sub-SMC grid, a higher effective temperature is required to achieve the same ionizing photon output. Specifically, stars must be hotter by $\sim$$\SI{2}{kK}$ to reach comparable $\log Q_\mathrm{H}$ values. This shift mirrors the temperature offset observed in the spectral classification, such that even at sub-SMC metallicity, stars of roughly spectral type O6 are required to efficiently ionize their surrounding hydrogen. This trend is in agreement with the work on stars in Sextans A from \citet{lor1:25}.

        For \HeI{}-ionizing radiation, models with $\sim$$\SI{23}{kK}$ ($\gtrsim$B0.7) already produce significant photon fluxes exceeding $\log(Q_\HeI{}/\mathrm{s^{-1}})>44$. However, only the most massive and hottest stars in our model grids, those with $\sim$$\SI{45}{kK}$ ($\gtrsim$O3.5), emit $\log(Q_\HeI{}/\mathrm{s^{-1}})>49$, which is sufficient to singly ionize helium in their surrounding medium. At lower metallicity, stars are generally hotter and more compact at a given mass, which increases the number of models exceeding this threshold. For \HeII{}-ionizing radiation, only the hottest models in our grid, namely those with temperatures above $\sim$$\SI{45}{kK}$ ($\gtrsim$O3.5), produce a non-negligible photon flux exceeding $\log(Q_\HeII{}/\mathrm{s^{-1}})>44$. At lower metallicity, where stellar winds are weaker the parameter space where stars emit hard ionizing photons extends to lower $T$ and $\log g$, underlining the role of massive stars as key cosmic engines in the Early Universe. Note that hot ($T\approx\SI{100}{kK}$) intermediate stripped star typically produce between $\log(Q_\HeII{}/\mathrm{s^{-1}})=\SIrange{40}{44}{}$ depending on their mass and mass-loss rate \citep{goe1:17}, and that low-metallicity early-type WRs can reach even $\log(Q_\HeII{}/\mathrm{s^{-1}})=\sim48$ \citep{gon1:25,san1:26}. Thus, the ionizing flux of a single early O-type star can produce up to two orders of magnitude more \HeII{}-ionizing radiation.
        
    \section{Summary and conclusions}
    \label{sec:conclusions}

        In this work, EWs and typical line ratios of key temperature diagnostic lines of OB stars, calculated from large grids of stellar atmosphere models spanning a wide metallicity range from $Z=1/31\,\zsun\,\text{--}\,1\,\zsun$, are presented. Each grid contains about $250$ stellar atmosphere models specified by temperature, surface gravity, and luminosity, whereas the latter is derived from interpolation from detailed stellar evolution tracks. The grids span a wide range of temperatures ($T=\SIrange{15}{62}{kK}$) and surface gravities (${\log(g/(\mathrm{cm\,s^{-2}}))=\numrange{4.4}{2.0}{}}$). All models are publicly accessible via the PoWR website.        
        
        We utilized the stellar atmosphere grids to study the changes in the EW of key temperature diagnostic lines as a function of metallicity. \HeI{} and \HeII{} lines are well-established temperature diagnostics that are present in all of our grids. We can see that with increasing metallicity, the He diagnostic lines shift in temperature due to the effect of back-warming in the atmosphere. While in our lowest metallicity grid, \HeI{} lines can still be present in the spectra of stars as hot as $T=\SI{55}{kK}$, these lines are only present up to $T\lesssim\SI{48}{kK}$ in the Galactic grid. Given that at low metallicity metal lines become significantly weaker, classification schemes for the earliest O-type stars based on N lines \citep{wal1:04} fail. In this context, the persistence of \HeI{} lines at such high temperatures provides a promising alternative diagnostic for the classification of low-metallicity stars.

        For cooler stars, which lack \HeII{} lines, often Al, Mg, and Si lines are employed as additional temperature diagnostics. Our models have shown that the \AlIII{}\,$\lambda\,4529$ line is only visible in the spectra of Galactic stars, while \MgII{}\,$\lambda\,4481$ and \SiII{}\,$\lambda\lambda\,4128,4131$ are present in the coolest stars of our grids except at sub-SMC metallicity. Only the \SiIII{}\,$\lambda4553$ and the \SiIV{}\,$\lambda4089$ lines can be seen in all of our grids. These lines appear in our models with temperatures below $T\lesssim\SI{35}{kK}$ and are thus a good temperature diagnostic for most B-type stars.

        We further demonstrated that various \HeI{} lines respond differently to temperature and surface gravity. Their EWs peak at distinct temperatures. Fitting several \HeI{} lines in the optical range allows at least for rough constraints on the temperature of cool metal-poor stars even when other metal lines are not detectable.

        Using stellar templates of O- and B-type stars from our Galaxy, we established quantitative boundaries of typically employed line ratios for the spectral type classification of massive stars. We have demonstrated that at solar metallicity these classification boundaries barely overlap, allowing for a consistent classification. For O-type stars the classification using ratios of different ionization levels of helium shows some correlation of surface gravity to the assigned spectral type, leading to a certain temperature spread. The line ratios employed for the late-O and B-type stars based on metal lines are  mostly insensitive to surface gravity, but the corresponding metal lines become undetectable at low metallicity. Only detailed stellar atmosphere modeling and consideration of the different temperature sensitivities of the \HeI{} lines will enable an accurate temperature estimation in this case.

        Lastly, we give the hydrogen and helium ionizing fluxes as predicted from our stellar atmosphere models in the $T$-$\log g$ parameter space. We have demonstrated that stars at lower metallicity tend to emit more hard ionizing radiation due to the weakening of their stellar wind, which becomes transparent for these wavelengths. This underlines the importance of massive stars as key feedback agents in the early Universe.    

    \section{Data availability}

        All grid models presented in this paper can be accessed via the user-friendly PoWR web-interface via \url{www.astro.physik.uni-potsdam.de/PoWR/} in the ``OB model grids'' category. The new grids presented in this work have the suffix ``-Vd3''. For the reproducibility of this work and long-term storage, we uploaded the normalized UV and optical spectra used in this paper on Zenodo and can be accessed via \url{https://doi.org/10.5281/zenodo.18173695}. For scientific purposes, we strongly advise using the model spectra from the PoWR homepage.

    \begin{acknowledgements}
        We thank the anonymous referee for their suggestions that significantly improved this paper.
        This work was performed in part at the Aspen Center for Physics, which is supported by National Science Foundation grant PHY-2210452.
        The authors thank Prof. Dr. Huirong Yan (Potsdam University) for providing additional computing resources. DP acknowledges financial support from the FWO in the form of a junior postdoctoral fellowship No. 1256225N. SRS acknowledges financial support by the Deutsches Zentrum f\"ur Luft und Raumfahrt (DLR) grants 50OR2108. AACS and VR acknowledge support by the Deutsche Forschungsgemeinschaft (DFG, German Research Foundation) in the form of an Emmy Noether Research Group -- Project-ID 445674056 (SA4064/1-1, PI Sander). This project was co-funded by the European Union (Project 101183150 - OCEANS).
        This work was facilitated by the International Space Science Institute (ISSI) in Bern, through ISSI International Team project 512 (Multiwavelength View on Massive Stars in the Era of Multimessenger Astronomy, PI Oskinova).	TS acknowledges support from the Israel Science Foundation (ISF) under grant number 0603225041 and from the European Research Council (ERC) under the European Union's Horizon 2020 research and innovation program (grant agreement 101164755/METAL).
    \end{acknowledgements}
 		
 	\bibliographystyle{aa}                                                         
 	\bibliography{astro} 
 		
 \clearpage        
 	
    \begin{appendix}
    \section{Interpolating between stellar evolution tracks}
    \label{app:interpol}

\begin{figure}[th]
            \centering
            \includegraphics[trim={0cm 8.5cm 0cm 28cm},clip, width=0.48\textwidth]{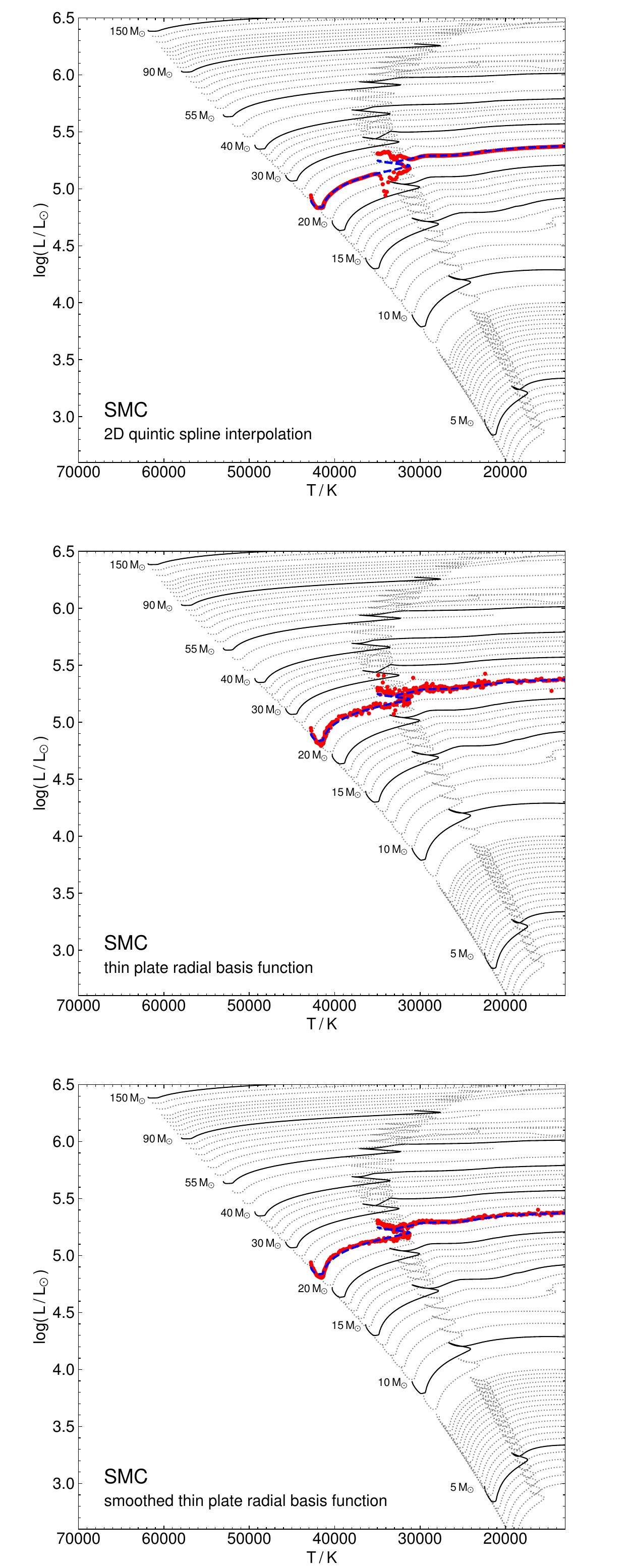}
            \caption{HRD demonstrating the efficiency of different interpolation methods in reproducing the $24\,\msun$ stellar evolution track of the SMC models. Solid black lines and dotted gray lines are the tracks used for the interpolation. The red dots are the interpolated luminosity values for the temperature and surface gravity combinations of the $24\,\msun$ stellar evolution track. The dashed blue line shows the actual (excluded) $24\,\msun$ stellar evolution track.}
            \label{fig:interpol}
        \end{figure}

        Interpolating between two or multiple evolutionary tracks is a nontrivial challenge. In this study, our objective is to derive the luminosity for a stellar atmosphere model based on its temperature and surface gravity. However, stellar evolution tracks are not a simple function of temperature and surface gravity as the tracks go back and forth in the HRD or the Kiel diagram (e.g., the terminal-age main-sequence (TAMS) hook). To ensure the reliability of our interpolated luminosity values, we performed extensive testing on various interpolation methods. During this process, one of the stellar evolution tracks was systematically excluded, and the interpolation was carried out using the remaining tracks. The temperature and surface gravity values of the excluded track were then employed to calculate the interpolated luminosity values. Subsequently, these values were compared against the actual luminosity values of the excluded track. In Fig.~\ref{fig:interpol}, a few selected examples of different interpolation methods are displayed.

        Our first approach was to perform a two-dimensional spline interpolation using linear, cubic, and quintic splines. Even though the results improved with more complex functions, all of them were unable to resolve the area around the TAMS (see first panel of Fig.~\ref{fig:interpol}). The differences between the interpolated values and the actual track were larger than ${\Delta\log(L/\lsun)>0.3}$, which would have a drastic impact on the spectral appearance of a stellar atmosphere model.

        As a second approach, we conducted interpolations with radial basis functions, a more advanced interpolation method that is typically used for high-dimensional complex functions. Again, we tested various basis functions, such as linear, cubic, Gaussian, and multiquadric. However, the best results were achieved when using a radial basis function interpolation with a thin-plate spline basis function. The interpolated track using this technique is shown in the second panel of Fig.~\ref{fig:interpol}. It is evident that the radial basis function can mimic the stellar evolution track fairly well, but still shows some scatter around the TAMS, which is on the order of ${\Delta\log(L/\lsun)\approx0.1}$. 

        To improve the reliability of the interpolated values obtained through the radial basis function, we computed luminosity values for 100 combinations of temperature and surface gravity within the range of ${\Delta T_\mathrm{interpol}=\pm\SI{100}{K}}$ and ${\Delta\log(g_\mathrm{interpol}/(\mathrm{cm\,s^{-2}}))=\pm0.02}$. From these 100 luminosity values, the average was computed, resulting in improved agreements between the interpolated values and the actual stellar evolution track (see third panel of Fig.~\ref{fig:interpol}). This approach not only accounts for the variation within the chosen parameter ranges but also enhances the robustness of the interpolation results.
        
    \clearpage
    \newpage
    \onecolumn
    \section{Additional figures about the grid models}

        \begin{figure*}[th]
            \centering
            \includegraphics[trim={3.75cm 3.5cm 4cm 4cm},clip, width=0.95\textwidth]{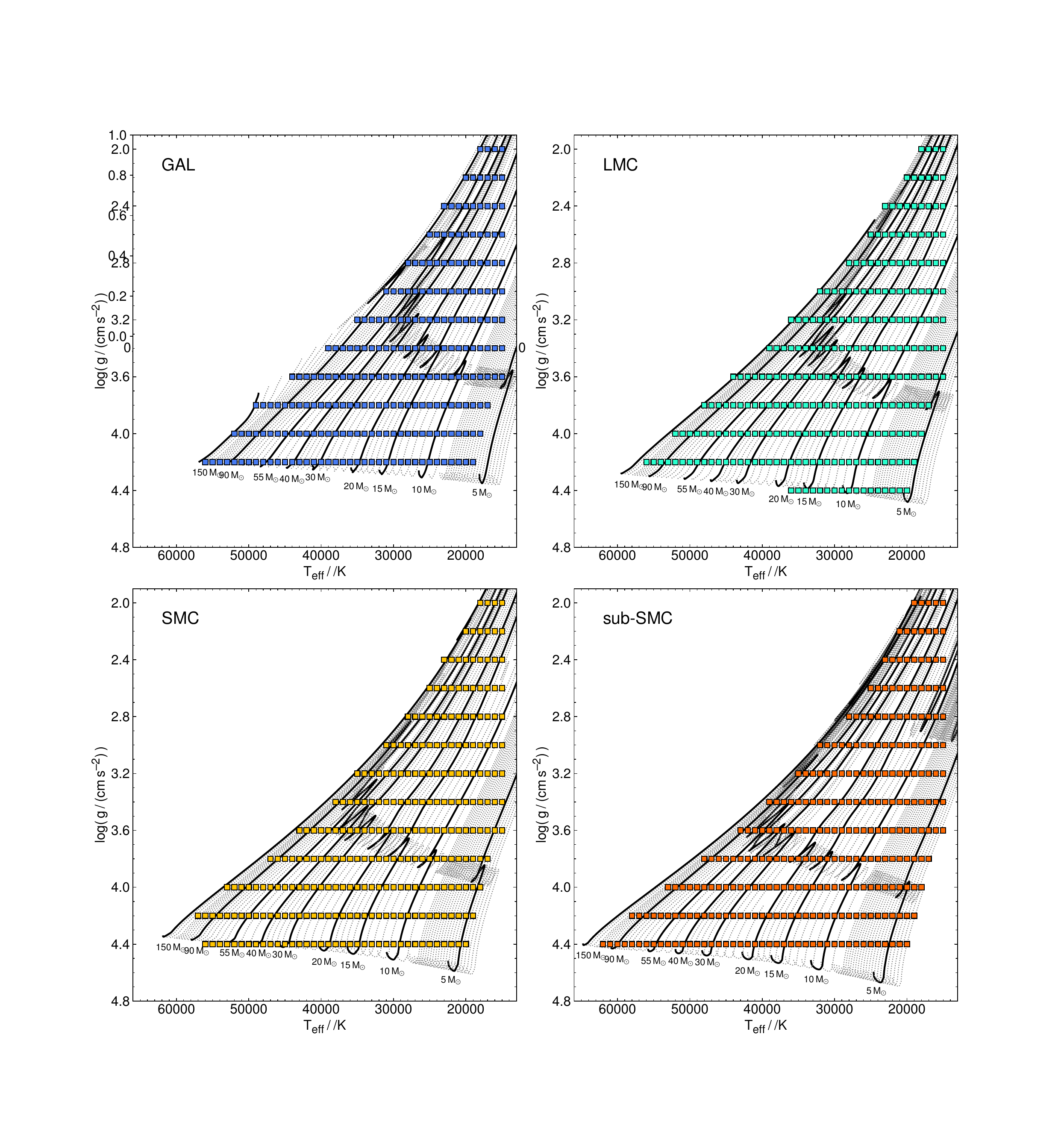}
            \caption{Kiel diagrams showing the MIST evolutionary tracks (lines) over-plotted with the sets of temperatures and surface gravities covered by our new stellar atmosphere models (squares). We show here the $T_\mathrm{eff}$ from the MESA models and the $T_\ast$ of the stellar atmosphere models. Note that for OB stars, which have optically thin winds, the difference between $T_\mathrm{eff}$ and $T_\ast$ is very small ($\lesssim1\%$). We show all MIST tracks available for the initial mass range of $M=\numrange{4}{150}\,\msun$ as lines. For clarity we show only the tracks with initial masses $5\,\msun$, $10\,\msun$, $15\,\msun$, $20\,\msun$, $30\,\msun$, $40\,\msun$, $55\,\msun$, $90\,\msun$, and $150\,\msun$ as solid black lines and the remaining tracks as dotted gray lines.}
            \label{fig:kiel}
        \end{figure*}
    \clearpage
    \newpage
    \onecolumn
    \section{Spectral ranges used for the different spectral lines}
    
    \begin{figure*}[th]
        \centering
        \begin{minipage}{0.48\textwidth}
            \centering
            \captionof{table}{List of the spectral ranges used to calculate the EWs of the phosphoric lines.}
            \begin{tabular}{lc}\hline \hline \rule{0cm}{2.8ex}%
                \rule{0cm}{2.2ex} Line & Spectral Range \\
                \rule{0cm}{2.2ex} & $[\AA]$ \\
                \hline \rule{0cm}{3.4ex}%
                \rule{0cm}{2.8ex} \HeI{}\,$\lambda\,4120$   & $\numrange{4118}{4125}$ \\
                \rule{0cm}{2.8ex} \HeI{}\,$\lambda\,4471$   & $\numrange{4468}{4475}$ \\
                \rule{0cm}{2.8ex} \HeI{}\,$\lambda\,5876$   & $\numrange{5871}{5881}$ \\
                \rule{0cm}{2.8ex} \HeI{}\,$\lambda\,7065$   & $\numrange{7063}{7069}$ \\
                \rule{0cm}{2.8ex} \HeII{}\,$\lambda\,4541$  & $\numrange{4530}{4550}$ \\
                \rule{0cm}{2.8ex} \HeII{}\,$\lambda\,5412$  & $\numrange{5397}{5427}$ \\
                \rule{0cm}{2.8ex} \MgII{}\,$\lambda\,4481$  & $\numrange{4479.5}{4484}$ \\
                \rule{0cm}{2.8ex} \AlIII{}\,$\lambda\,4529$   & $\numrange{4526}{4532}$ \\
                \rule{0cm}{2.8ex} \SiII{}\,$\lambda\lambda\,4128, 4131$  & $\numrange{4126}{4134}$ \\
                \rule{0cm}{2.8ex} \SiIII{}\,$\lambda\,4553$   & $\numrange{4550}{4556}$ \\
                \rule{0cm}{2.8ex} \SiIV{}\,$\lambda\,4089$   & $\numrange{4086}{4091}$ \\
                \rule{0cm}{2.8ex} \Hbeta{}                  & $\numrange{4841}{4881}$ \\
                \hline
            \end{tabular}
            \rule{0cm}{2.8ex}%
            \begin{minipage}{0.95\linewidth}
                \ignorespaces 
            \end{minipage}
            \label{tab:EW}
        \end{minipage}
        \hspace{2ex}
        \begin{minipage}{0.48\textwidth}
            \centering
            \includegraphics[trim={1.3cm 1.2cm 0.7cm 1.7cm},clip, width=\textwidth]{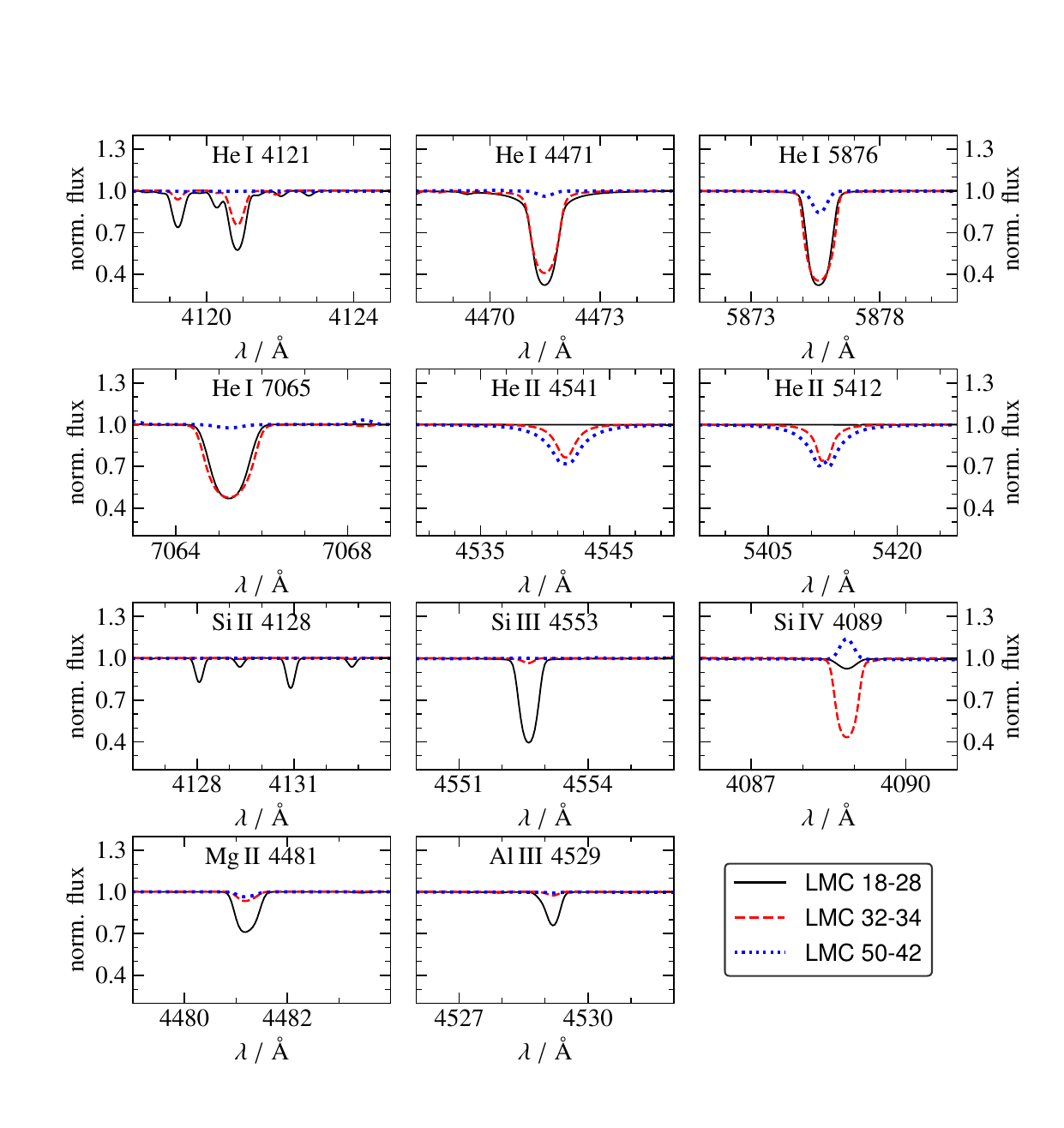}
            \caption{Comparison of all considered temperature and surface gravity diagnostic lines for three models of the LMC grid.}
            \label{fig:all_lines}
        \end{minipage}

    \end{figure*}

    \section{Line ratios employed for spectral classification}    
    \label{app:temp}
    
        \begin{figure*}[thb]
            \centering
            \includegraphics[trim={2.5cm 0.5cm 3.5cm 2.4cm},clip,width=0.95\textwidth]{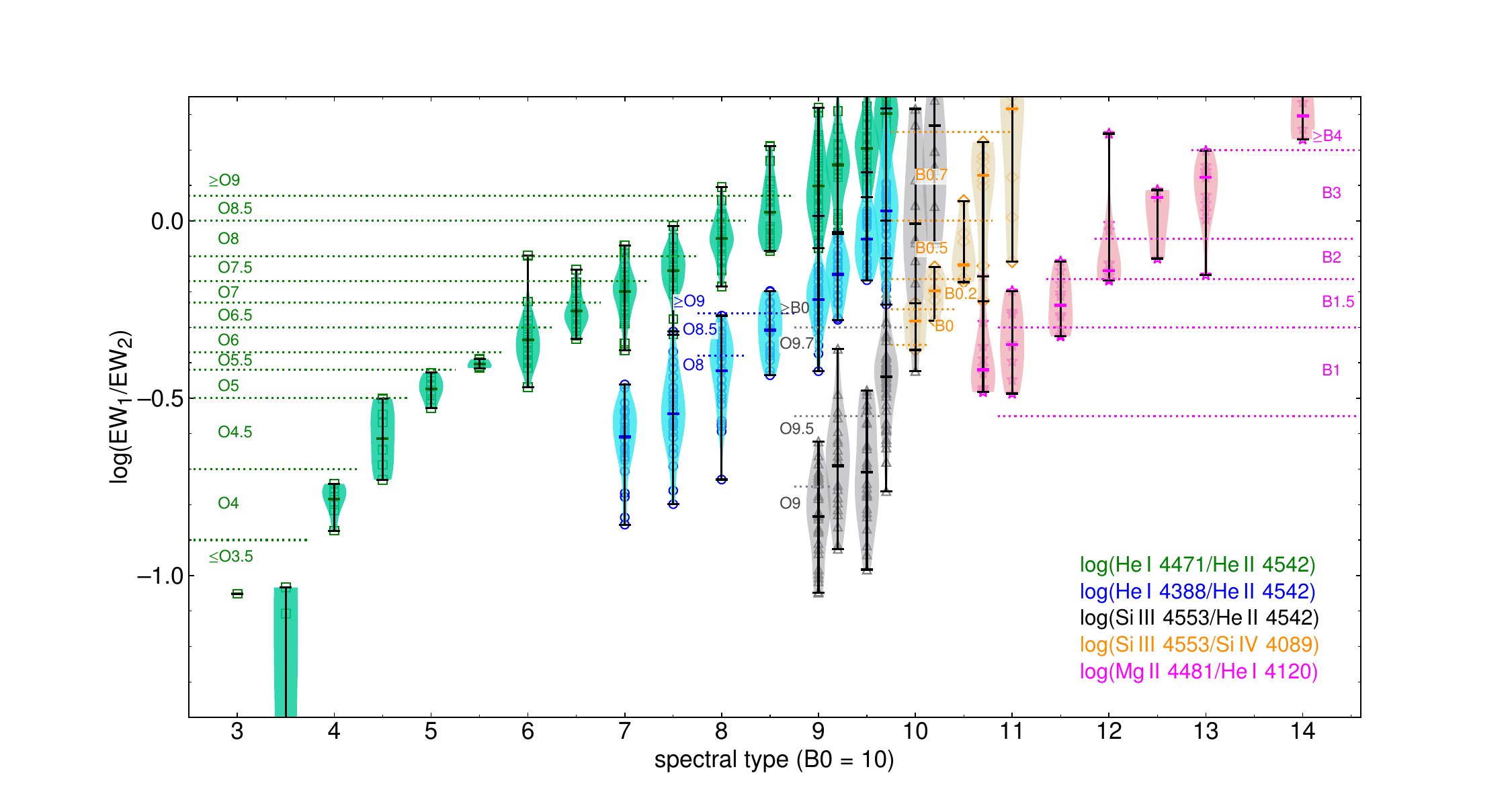}
            \caption{EW line ratios of selected helium and metal lines typically used for spectral classification as a function of spectral type. Colored markers represent measured EW from individual OB template stars \citep{mai1:11,mai1:13,mai1:16,neg1:24}. Violin plots illustrate the distribution of EWs for each spectral type, with the bold markers indicating the medians. Dotted lines denote the assigned classification boundaries adopted in this work for each spectral type (see Table~\ref{tab:ratios}).}
            \label{fig:obs_ratios}
        \end{figure*}

        \begin{table*}[htbp]
            \centering
            \caption{Criteria adopted for the spectral classification of our models, calibrated using Galactic OB star templates (see Fig.~\ref{fig:obs_ratios}).}
            \begin{tabular}{lll}\hline \hline \rule{0cm}{2.8ex}%
                \rule{0cm}{2.2ex} spectral type & \multicolumn{1}{c}{main criterion}  & \multicolumn{1}{c}{additional criterion}\\
                \hline \rule{0cm}{3.4ex}%
                \rule{0cm}{2.8ex} >O3.5 & $-0.90>\log(\HeI{}\,\lambda4471/\HeII{}\,\lambda4542)$&\\
                \rule{0cm}{2.8ex} \phantom{>}O4 & $-0.70\geq\log(\HeI{}\,\lambda4471/\HeII{}\,\lambda4542)> -0.90$&\\
                \rule{0cm}{2.8ex} \phantom{>}O4.5 & $-0.50\geq\log(\HeI{}\,\lambda4471/\HeII{}\,\lambda4542)> -0.70$&\\
                \rule{0cm}{2.8ex} \phantom{>}O5 & $-0.50\geq\log(\HeI{}\,\lambda4471/\HeII{}\,\lambda4542)> -0.42$&\\
                \rule{0cm}{2.8ex} \phantom{>}O5.5 & $-0.42\geq\log(\HeI{}\,\lambda4471/\HeII{}\,\lambda4542)> -0.37$&\\
                \rule{0cm}{2.8ex} \phantom{>}O6 & $-0.37\geq\log(\HeI{}\,\lambda4471/\HeII{}\,\lambda4542)> -0.30$&\\
                \rule{0cm}{2.8ex} \phantom{>}O6.5 & $-0.30\geq\log(\HeI{}\,\lambda4471/\HeII{}\,\lambda4542)> -0.23$&\\
                \rule{0cm}{2.8ex} \phantom{>}O7 & $-0.23\geq\log(\HeI{}\,\lambda4471/\HeII{}\,\lambda4542)> -0.17$&\\
                \rule{0cm}{2.8ex} \phantom{>}O7.5 & $-0.17\geq\log(\HeI{}\,\lambda4471/\HeII{}\,\lambda4542)> -0.10$&\\
                \rule{0cm}{2.8ex} \phantom{>}O8 & $-0.38\geq\log(\HeI{}\,\lambda4388/\HeII{}\,\lambda4542)$ & $-0.10\geq\log(\HeI{}\,\lambda4471/\HeII{}\,\lambda4542)> 0.00$\\
                \rule{0cm}{2.8ex} \phantom{>}O8.5 & $-0.24\geq\log(\HeI{}\,\lambda4388/\HeII{}\,\lambda4542)>-0.38$ & $\phantom{-}0.00\geq\log(\HeI{}\,\lambda4471/\HeII{}\,\lambda4542)> 0.07$\\
                \rule{0cm}{2.8ex} \phantom{>}O9 & $-0.70\geq\log(\SiIII{}\,\lambda4553/\HeII{}\,\lambda4542)$ & $-0.24\geq\log(\HeI{}\,\lambda4388/\HeII{}\,\lambda4542)$\\
                \rule{0cm}{2.8ex} \phantom{>}O9.5 & $-0.45\geq\log(\SiIII{}\,\lambda4553/\HeII{}\,\lambda4542)>-0.70$ & \\
                \rule{0cm}{2.8ex} \phantom{>}O9.7 & $-0.20\geq\log(\SiIII{}\,\lambda4553/\HeII{}\,\lambda4542)>-0.45$ & \\
                \rule{0cm}{2.8ex} \phantom{>}B0 & $-0.25\geq\log(\SiIII{}\,\lambda4553/\SiIV{}\,\lambda4089)>-0.35$ &  $-0.55\geq\log(\MgII{}\,\lambda4481/\HeI{}\,\lambda4120)$\\
                \rule{0cm}{2.8ex} \phantom{>}B0.2 & $-0.165\geq\log(\SiIII{}\,\lambda4553/\SiIV{}\,\lambda4089)>-0.25$ &  $-0.55\geq\log(\MgII{}\,\lambda4481/\HeI{}\,\lambda4120)$\\
                \rule{0cm}{2.8ex} \phantom{>}B0.5 & $\phantom{-}0.00\geq\log(\SiIII{}\,\lambda4553/\SiIV{}\,\lambda4089)>-0.165$ &  $-0.55\geq\log(\MgII{}\,\lambda4481/\HeI{}\,\lambda4120)$\\       
                \rule{0cm}{2.8ex} \phantom{>}B0.7 & $\phantom{-}0.25\geq\log(\SiIII{}\,\lambda4553/\SiIV{}\,\lambda4089)>\phantom{-}0.00$ &  $-0.55\geq\log(\MgII{}\,\lambda4481/\HeI{}\,\lambda4120)$\\                
                \rule{0cm}{2.8ex} \phantom{>}B1 &  $-0.30\geq\log(\MgII{}\,\lambda4481/\HeI{}\,\lambda4120)>-0.55$&\\                
                \rule{0cm}{2.8ex} \phantom{>}B1.5 &  $-0.165\geq\log(\MgII{}\,\lambda4481/\HeI{}\,\lambda4120)>-0.30$&\\        
                \rule{0cm}{2.8ex} \phantom{>}B2 &  $-0.05\geq\log(\MgII{}\,\lambda4481/\HeI{}\,\lambda4120)>-0.165$&\\                         
                \rule{0cm}{2.8ex} \phantom{>}B3 &  $\phantom{-}0.20\geq\log(\MgII{}\,\lambda4481/\HeI{}\,\lambda4120)>-0.05$&\\
                \rule{0cm}{2.8ex} BHG & \multicolumn{1}{c}{\Hbeta{}\,$<0$} & \\  
                \hline
            \end{tabular}
            \rule{0cm}{2.8ex}%
            \begin{minipage}{0.95\linewidth}
                \ignorespaces 
            \end{minipage}
            \label{tab:ratios}
        \end{table*}

\end{appendix}
\end{document}